\pdfoutput=1
\documentclass[aps, reprint, superscriptaddress, nofootinbib, showpacs, floatfix]{revtex4-1}
\usepackage{amsmath, latexsym, amssymb, hyperref, graphicx, color, multirow}
\usepackage[table]{xcolor}
\usepackage{tabularx}
\usepackage[T1]{fontenc}
\usepackage[capitalize]{cleveref}
\definecolor{nicered}{rgb}{.7,.1,.1}
\definecolor{nicegreen}{rgb}{.1,.5,.1}
\definecolor{darkblue}{rgb}{0,0,.5}
\hypersetup{colorlinks, citecolor=nicegreen,linkcolor=nicered, urlcolor=darkblue}

\begin{document}


\title{Higgs Sector of the Left-Right Symmetric Theory}
\author{Alessio Maiezza}
\email{alessiomaiezza@gmail.com}
\affiliation{International Centre for Theoretical Physics, Trieste, Italy}
\author{Goran Senjanovi\'c}
\email{goran@ictp.it}
\affiliation{International Centre for Theoretical Physics, Trieste, Italy}
\affiliation{Gran Sasso Science Institute, L'Aquila, Italy }
\author{Juan Carlos Vasquez}
\affiliation{ Centro Cient\'ifico Tecnol\'ogico de Valpara\'iso-CCTVal, \\
Universidad T\'ecnica Federico Santa Mar\'ia, Valpara\'iso, Chile}
\email{juan.vasquezcar@usm.cl}
\affiliation{International Centre for Theoretical Physics, Trieste, Italy}
\affiliation{International School for Advanced Studies/INFN, Trieste, Italy}


\begin{abstract}
\noindent
We perform an in-depth analysis of the Higgs sector in the Minimal Left-Right Symmetric Model and compute the scalar mass spectrum and associated mixings, offering simple physical and symmetry arguments in support of our findings. We identify the tree-level quartic and cubic potential couplings in terms of the physical states and compute the quantum corrections for the latter ones. The deviations from the Standard Model prediction of the cubic Higgs doublet coupling are considered.
Moreover we discuss the possible implications concerning the stability of the potential under the renormalization-group-equations evolution. In particular we examine three possible energy scales of parity restoration: LHC reach, next hadronic collider and very high energy relevant for grand unification.
\end{abstract}

\maketitle

\section{Introduction}

There has been a great deal of interest in the Left-Right symmetric electro-weak gauge theory~\cite{leftright, Senjanovic:1978ev}
in recent years due its potential accessibility at the LHC. After more than four decades since its birth, there is finally hope that experiment could confirm it. Moreover, it has emerged~\cite{Nemevsek:2012iq} that the minimal such model is a self-contained and predictive theory of neutrino mass in full analogy with the standard model (SM) for the Higgs origin of charged fermions masses. We can say that what seemed originally its curse, the prediction of massive neutrino, over the years turned into a great blessing. In this, the crucial role was played by the seesaw mechanism~\cite{Minkowski:1977sc, Mohapatra:1979ia, seesaw} which not only suggestively accounts for small neutrino mass, but moreover makes it be of Majorana nature. This implies Lepton Number Violation (LNV) both at low energies through the neutrinoless double beta decay~\cite{Racah:1937qq} and at high
energies through a production of same sign charged lepton pairs at hadronic colliders~\cite{Keung:1983uu}. In the minimal Left-Right symmetric model (LRSM) there is a deep connection between these processes~\cite{Tello:2010am}.

There has recently been another important advancement in the minimal LR model, the analytic expression for the right-handed quark mixing matrix, in all of the parameter space~\cite{Senjanovic:2014pva}. It showed that the left and right-handed mixing angles are remarkably close to each other in spite of near maximal parity violation in low energy weak interactions.

The LR symmetric theory is the simplest realization of the idea of the restoration of parity at the fundamental level. LR symmetry is broken spontaneously, and parity violation is supposed to be a low energy accident. Since it was known fairly early that the right-handed (RH) charged gauge boson $W_R$ had to be very heavy due to its impact on the $K_L- K_S$ mass difference, on the order of few TeV~\cite{Beall:1981ze, Mohapatra:1983ae, Ecker:1985vv}, one had to wait for the advent of LHC in order to study it experimentally. This limit has been revisited in recent years~\cite{Zhang:2007da} and definitively estimated to lie in the full LHC reach~\cite{Maiezza:2010ic, Bertolini:2014sua}, which ranges up to $\sim6$ TeV for the $W_R$ mass~\cite{Ferrari:2000sp}. This value would make neutrinoless double decay likely to be seen, even if it were not due to neutrino mass. The LHC is slowly but surely getting there~\cite{Khachatryan:2014dka}, with the limit $M_{W_R} \gtrsim 3$ TeV in a large portion of the parameter space of RH neutrino masses.

It is then important to study carefully the LRSM in its full glory, including the Higgs sector.
The original analysis of the Higgs sector goes back almost forty years~\cite{Senjanovic:1978ev}, and it had cleared some essential features of the LR theory, such as the issue of flavor violation in the neutral scalar sector.
It was quite comprehensive, but it had to do with the outdated version of the theory with Dirac neutrinos. The changes are not dramatic, basically they reduce to the existence of doubly charged scalars. They are important though to be taken into account and were discussed first in~\cite{Deshpande:1990ip, Duka:1999uc, Kiers:2005gh, Barenboim:2001vu} and most recently in~\cite{Tello:2012qda, Maiezza:2016bzp, Dev:2016dja}.

The previous studies lacked the computation of the masses and mixings of scalar particles in the whole (phenomenologically relevant) parameter space. It is not the only reason that drove us to go through this not very inspiring task plagued by computational tedium, although we believe that this by itself ought to suffice. The main issue for the low scale LRSM, the one accessible at the LHC, is the issue of stability and perturbativity of the potential at higher energies.
Namely, the low energy constraints from meson mixing, the same that drive $W_R$ to be heavy, imply a stringent limit~\cite{Ecker:1985vv, Maiezza:2010ic, Blanke:2011ry} on the mass of the additional Higgs doublet necessarily present in the minimal model on the order of 20 TeV~\cite{Bertolini:2014sua}, which leads to a worry of possibly too large couplings in the Higgs potential. This was recently studied in~\cite{Maiezza:2016bzp}, where it was deduced that the theory can be perturbative for the LHC scale of symmetry breaking, but that it lives dangerously. We discuss this issue further, and in particular address the question of the cut-off scale where the theory ceases to work. We show that the closeness of the cut-off to the LR symmetry breaking scale brings important consequences on the parametric space of the model.

There is more to it. After all, the LR scale is not predicted by theory and strictly speaking it can be anywhere between TeV and the Planck scale. Obviously, the LHC reach is of great importance but one should be getting ready for future hadronic colliders, now being planned. There have already been studies devoted to this possibility, such as~\cite{Dev:2016dja}, with a hope of reaching the LR scale around 20 TeV. We find that in this case the theory is perfectly perturbative and the cut-off can be far from the mass of $W_R$, allowing for a natural suppression of ultra-violet (UV) non-renormalizable operators.

Another important scale is the one suggested by the minimal $SO(10)$ grand unified theory (GUT), around $10^{10}$ GeV. We run the whole parametric space of the model in order to check if the scalar sector remains perturbative preserving the picture of the two step symmetry breaking. As a result, we get the generic constraint that the quartic couplings have to be of order of few percent, favoring marginally light scalars.

The rest of the paper is organized as follows. In the next section we review briefly the essential features of the minimal Left-Right symmetric model. In the section~\ref{III} we give the scalar mass spectrum and the relevant mixings. We offer simple symmetry arguments behind our results in order to facilitate the reading of the paper and as a check of our computations. We also give the physical quartic and cubic couplings and discuss the deviations from the SM results. In the section~\ref{IV} we apply our results to the question of stability and perturbativity of the potential. We pay special attention to the issue of the cut-off which signals the breakdown of perturbativity at higher energies. In section~\ref{Qver} we consider the vertex renormalization, explicitly showing where the relevant vertices vary from the tree-level ones. Finally, in section~\ref{V} we offer a summary and outlook of our results. The paper is completed by three Appendices where we give some of the more technical details.

\section{Salient features of the minimal Left-Right Symmetric Model}

\paragraph*{Gauge group and field content.} The minimal LR symmetric theory is based on the $\mathcal{G}_{LR}\equiv SU(2)_L \times SU(2)_R \times U(1)_{B-L}$
gauge group (suppressing color) and a symmetry between the left and the right sectors.
Quarks and leptons come in LR symmetric representations
\begin{equation}
  Q_{L,R } = \left( \begin{array}{c} u \\ d \end{array}\right)_{L,R},\qquad
  \ell_{L,R} = \left( \begin{array}{c} \nu \\ e \end{array}\right)_{L,R}.
  \label{ds21}
\end{equation}
The formula for the electromagnetic charge becomes~\cite{Mohapatra:1980qe} $Q_{em} = I_{3 L} + I_{3 R} + {B - L \over 2}$
which trades the hard to recall hyper-charge of the SM for $B-L$, the physical anomaly-free global symmetry of the SM, now gauged. Both LR symmetry and the gauged $B-L$ require the presence of RH neutrinos.

The Higgs sector consists of the following multiplets~\cite{Minkowski:1977sc,Mohapatra:1979ia,Mohapatra:1980yp}: the bi-doublet
$\Phi \in (2_L,2_R,0)$ and the $SU(2)_{L,R}$ triplets $\Delta_L \in (3_L,1_R,2)$ and $\Delta_R \in
(1_L,3_R,2)$
\begin{equation}
  \Phi = \left[\begin{array}{cc}\phi_1^0&\phi_2^+\\\phi_1^-&-\phi_2^{0*}\end{array}\right]\,,
  \qquad
  \Delta_{L, R} = \left[ \begin{array}{cc} \delta^+ /\sqrt{2}& \delta^{++} \\
  \delta^0 & -\delta^{+}/\sqrt{2} \end{array} \right]_{L,R}.
  \label{ds32}
\end{equation}
We denote the bi-doublet as two SM model $Y= -1$ doublets $\phi_i$ as in~\cite{Tello:2012qda}
\begin{equation}
\Phi = \left[\phi_1, \, \epsilon \phi_2^*\right];\,\,\,\,\,\,\, \phi_i = \left(\begin{array}{c}\phi_i^0\\ \phi_i^-\end{array}\right),\,i=1,2\,,
  \qquad
   \label{phi}
\end{equation}
with $\epsilon=i\sigma_2$.
This manifest SM notation allows one to make direct comparison between the LR theory and the SM with two Higgs doublets.

\paragraph*{Symmetry breaking.}The symmetry breaking proceeds through two steps. First, at high scale with the breaking of
$SU(2)_R \times U(1)_{B-L} \to U(1)_Y $ through the vacuum expectation values (vev)~\cite{Mohapatra:1980yp}
\begin{equation}
\langle \delta_R^0 \rangle \equiv v_R, \,\,\,  \langle \delta_L^0 \rangle \equiv v_L = 0\,,
\end{equation}
which is responsible for the masses of new gauge bosons $W_R$, $Z_R$
\begin{equation}\label{eq:WRmass}
M_{W_R}^2 \simeq g_R^2 v_R^2,\,\,\,\,M_{Z_R}^2 \simeq (2 g_R^2+g_{B-L}^2) v_R^2,
\end{equation}
where $g_R,g_{B-L}$ are the gauge couplings of the $SU(2)_R,U(1)_{B-L}$ groups. Moreover, $v_R$ gives large masses to RH neutrinos $\nu_R$, denoted $N$ hereafter.

Next, at low scale with the usual SM symmetry breaking through (from here on we use the notation $\sin \gamma=s_\gamma, \cos \gamma=c_\gamma, \tan \gamma=t_\gamma$ for any angle $\gamma$)
\begin{equation}
\langle \Phi \rangle = v\,  \text{diag} (c_\beta,- e^{-i a}s_\beta)\,,
\end{equation}
which gives the mass to the LH charged gauge boson $M_W^2 = g_L^2/2 \, v^2$.
In turn, this in general produces a small vev for the left-handed triplet $v_L$, with
$ v_L \propto v^2/v_R$~\cite{Mohapatra:1980yp}, ensuring the usual dominant doublet symmetry breaking of the SM symmetry. The oblique parameters impose a bound $v_L\lesssim 5$ GeV, however in the see-saw picture that we follow, this bound becomes much more stringent since $v_L$ directly contributes to neutrino mass.

\paragraph*{Parity restoration: $\mathcal{P}$ or $\mathcal{C}$.}The discrete LR symmetry can be shown to be either a generalized parity $\mathcal{P}$ or a generalized charge conjugation $\mathcal{C}$~\cite{Maiezza:2010ic}. Under these, the fields transform as follows
\begin{equation} \mathcal{P}:\left\{
	\begin{array}{c} f_L \leftrightarrow f_R \\
	\Phi \leftrightarrow \Phi^\dag \\
	\Delta_L \leftrightarrow \Delta_R
	\end{array} \right.
\qquad
\mathcal{C}:\left\{
	\begin{array}{c}
	f_L \leftrightarrow (f_R)^c \\
	\Phi \leftrightarrow \Phi^T \\
	\Delta_L \leftrightarrow \Delta_R^*
	\end{array}\right.
\label{eq:PCdef}
\end{equation}
where $(f_R)^c=C\gamma_0f_R^*$ is the usual charge-conjugate spinor. These symmetries imply $g_L=g_R \equiv g$ and strongly characterize the form
of the scalar potential that we are going to discuss.

\section{The Higgs scalar sector: masses, mixings and couplings}\label{III}

\subsection {The Higgs potential.}
The most general  potential consistent with the $\mathcal{G}_{LR}$ gauge group, without assuming any discrete LR symmetry, is given in~\cite{Dekens:2015csa}. It is too messy to be presented here. After all, if one does not believe in LR symmetry, why assume the existence of $\Delta_L$ if $\Delta_R$ suffices by itself? Let us focus instead on the the part of the potential containing only the bi-doublet, since it is quite instructive and will ease the reader's pain in facing the full potential. Its general form is given by
\begin{align}\label{Vphi}
	& \mathcal{V}_\Phi=  -\mu_\Phi^2 \text{Tr}(\Phi^{\dagger}\Phi )-\tilde \mu_\Phi^2 [\text{Tr}(\tilde{\Phi}^{\dagger}\Phi) +h.c. ] \nonumber \\
	&+\lambda_1 [\text{Tr}(\Phi^{\dagger}\Phi )]^2
	+ \lambda_2[e^{id_{2}}\text{Tr}^2(\tilde{\Phi}\Phi^{\dagger}) +h.c.] \nonumber \\
	&+ \lambda_3[\text{Tr}(\tilde{\Phi}\Phi^{\dagger})\text{Tr}(\tilde{\Phi}^{\dagger}\Phi) ]+\lambda_4 \text{Tr}(\Phi^{\dagger}\Phi )[e^{id_4}\text{Tr}(\tilde{\Phi}\Phi^{\dagger}) +h.c. ]\,, \nonumber \\	
	\end{align}
where $\tilde \Phi = \epsilon \Phi^* \epsilon= [\phi_2,\epsilon\phi_1^*]$ simply amounts for the interchange of the two $SU(2)_L$ doublets $\phi_1$ and $\phi_2$, yet another advantage of using the notation used in~\eqref{phi}. We have used the phase freedom of $\Phi$ to make the mass term $\tilde \mu_\Phi$ real. The potential has two real mass parameters and six real quartic couplings. It is instructive to compare it with the two Higgs doublet model case in $SU(2)_L \times U(1)$, where one has three real mass terms and ten real quartic couplings~\cite{Gunion:2002zf}. In spite of being much more restricted, the above potential still allows for a spontaneous violation of $CP$ as shown originally in~\cite{Senjanovic:1978ev}, however the generated phase would be too small due to the large mass of the second doublet, to be discussed below.

Clearly, the $SU(2)_R$ gauge symmetry plays an important role in restricting the number of parameters. We will see that the generalized charge conjugation as LR discrete symmetry makes no further restriction whatsoever on this part of the potential, as opposed to generalized parity that makes the couplings real. Of course, both of these LR symmetries connect the couplings of the LH and RH triplets $\Delta_{L,R}$ and simplify the potential considerably.

\paragraph*{Case $\mathcal{C}.$} We start with case of the generalized charge conjugation
$\mathcal{C}$ as the LR symmetry, since the case $\mathcal{P}$ is simply obtained in the limit of some vanishing phases (see below).
This further restricts the numbers of the parameters in the potential which now reads as~\cite{Tello:2012qda,Dekens:2014ina}
\begin{widetext}
\begin{align}\label{CV}
	& \mathcal{V}_C= -\mu_\Phi^2 \text{Tr}(\Phi^{\dagger}\Phi )-\tilde \mu_\Phi^2 [\text{Tr}(\tilde{\Phi}^{\dagger}\Phi) +h.c. ]- \mu_\Delta^2 [\text{Tr}(\Delta_L\Delta_L^{\dagger})+\text{Tr}(\Delta_R\Delta_R^{\dagger})]+\lambda_1 [\text{Tr}(\Phi^{\dagger}\Phi )]^2
	\nonumber \\
	&
	+ \lambda_2[e^{id_{2}}\text{Tr}^2(\tilde{\Phi}\Phi^{\dagger}) +h.c.]+ \lambda_3[\text{Tr}(\tilde{\Phi}\Phi^{\dagger})\text{Tr}(\tilde{\Phi}^{\dagger}\Phi) ]+\lambda_4 \text{Tr}(\Phi^{\dagger}\Phi )[e^{id_4}\text{Tr}(\tilde{\Phi}\Phi^{\dagger}) +h.c. ]+ [\rho_1\text{Tr}^2(\Delta_L\Delta_L^{\dagger})\nonumber \\
	&  +\rho_2 \text{Tr}(\Delta_L\Delta_L)\text{Tr}(\Delta_L^{\dagger}\Delta_L^{\dagger})+ L \to R]
	 +\rho_3 \text{Tr}(\Delta_L\Delta_L^{\dagger})\text{Tr}(\Delta_R\Delta_R^{\dagger}) +\rho_4 [e^{ir_4}\text{Tr}(\Delta_L^{\dagger}\Delta_L^{\dagger})\text{Tr}(\Delta_R\Delta_R)+h.c.] \nonumber \\
	& + \alpha_1  [\text{Tr}(\Phi^{\dagger}\Phi)+\alpha_2 (e^{ic}\text{Tr}(\tilde{\Phi}\Phi^{\dagger})+h.c.)][\text{Tr}(\Delta_L\Delta_L^{\dagger})+\text{Tr}(\Delta_R\Delta_R^{\dagger})] \nonumber
	+ \alpha_3 [\text{Tr}(\Phi\Phi^{\dagger}\Delta_L\Delta_L^{\dagger})+\text{Tr}(\Phi^{\dagger}\Phi\Delta_R\Delta_R^{\dagger})] \nonumber \\
	&
+[\beta_1 e^{ib_1}\text{Tr}(\Phi\Delta_R\Phi^{\dagger}\Delta_L^{\dagger}) +\beta_2 e^{ib_2}\text{Tr}(\tilde{\Phi}\Delta_R\Phi^{\dagger}\Delta_L^{\dagger})
	+\beta_3e^{ib_3}\text{Tr}(\Phi\Delta_R\tilde{\Phi}^{\dagger}\Delta_L^{\dagger})+h.c. \,\,(\beta_i= 0\,\, \text{in the seesaw picture})]
	\end{align}
\end{widetext}
The potential appears messy, simply because we have more than one same type couplings: the bi-doublet self-couplings $\lambda_i$, the triplet self-couplings $\rho_i$ and mixed couplings $\alpha_i$ and $\beta_i$. It turns out that in the seesaw limit the $\beta$ terms can be safely ignored as we discuss now.

What helps is the separation of the two scales of symmetry breaking, and the fact that for the physically interesting seesaw picture of neutrino mass one can safely ignore the small $v_L$. Namely, its contribution to neutrino mass matrix has the form~\cite{Mohapatra:1980yp} $M_\nu \propto v_L/v_R\, M_N$, where $M_N$ denotes the mass matrix of RH neutrinos $N$. Thus for a large portion portion of RH neutrino mass parameter space, $v_L$ must be quite small. For example, even in the case when $N$ are light and the lightest one provides warm dark matter~\cite{Bezrukov:2009th} with $m_N \simeq \text{keV}$, one has $v_L \lesssim 10^{-6} $ GeV which can be safely ignored in the analysis of the potential.
In the scenario where RH neutrinos can be actually seen at the colliders, $m_N \gtrsim 10 $ GeV, $v_L$ becomes completely negligible.

In what follows we thus work in the limit $v_L = 0$ (or equivalently $\beta = 0$). The question is whether it is technically natural. The positive answer was given already in the original work~\cite{Mohapatra:1980yp} but we go through it once again for the sake of completeness.
It is easy to see that $v_L$ is generated by the $\beta$ terms in the potential, and the smallness of $v_L$ is directly controlled by the smallness of $\beta_i$ couplings. It is equally easy to see that in the limit $\beta_i= 0$ there is more symmetry in the potential, e.g. $\Delta_L \to - \Delta_L$ which guarantees its stability to all orders in perturbation theory. This symmetry is broken only by the Yukawa couplings of $\Delta_L$, the same ones that lead to the seesaw picture since $\Delta_{L,R}$ have the same couplings because of the LR symmetry. In short, $v_L$ is naturally small in the technical sense, and in principle its effect can be sub-dominant to the usual seesaw contribution of RH neutrinos to neutrino mass.

This said, it is fair to admit that an extremely small $\beta$, as does the smallness of neutrino mass itself, points to the possible large LR-scale, which is natural in the context of the $SO(10)$ grand unified theory. Namely, in the minimal model one predicts $v_R \simeq 10 ^{10}$ GeV~\cite{delAguila:1980qag}. For this reason, we also include here a section dedicated to the $SO(10)$ embedding of the LRSM.

Since the LR-scale on the order of TeV is still perfectly acceptable, both theoretically and phenomenologically, one may wonder whether there is a more natural alternative to small $\beta$. Indeed, it is sometimes claimed that this can be achieved by decoupling $\Delta_L$ from the theory in order to have its vev small. We disagree with this for a number of reasons that we now go through.

First, unlike small protected dimensionless couplings, large scales are not technically natural because they bring in the usual hierarchy problem. Second, in order to decouple $\Delta_L$ in the context of the spontaneous symmetry breaking one needs to break the discrete LR symmetry at a large scale by the $\mathcal{G}_{LR}$ gauge singlet vev~\cite{Chang:1983fu}. Notice that keeping  $\Delta_R$ light while decoupling  $\Delta_L$ requires the usual fine-tuning between the original symmetric mass terms and the corrections induced by the singlet vev. Unlike in the case of small $\beta$, there is no protective symmetry here.  Moreover, a decoupled ad-hoc singlet is physically equivalent to the soft, non-spontaneous, breaking, and thus not well motivated.

If the LRSM is embedded in the $SO(10)$ theory however, the $\mathcal{G}_{LR}$ parity odd singlets are often automatically present~\cite{Chang:1983fu},  but then, as mentioned above, $M_{W_R}$ is predicted to be huge, around $10^{10}$ GeV~\cite{delAguila:1980qag}, and one is left basically with the SM at low energies (and massive neutrinos). One may find ways to lower $M_{W_R}$, but in that case one loses all the predictivity of grand unification.

Imagine for a moment that in any case one does invoke the GUT fields to argue in favor of a parity odd
$\mathcal{G}_{LR}$ gauge singlet field. In this case the LR theory has to remain perturbative and consistent all the way to the GUT scale. We will show in the following section that for the LR scale accessible at the LHC, the theory breaks down quite quickly. It helps to raise the LR scale to the one reachable at the future colliders, but it is still not enough, the quartic couplings become large well below the GUT scale.

Still, one can opt for the parity odd $\mathcal{G}_{LR}$ gauge singlet and claim that this helps the domain wall problem since the domain walls can be washed by the subsequent inflation. However, the domain wall problem is not so serious, for it may be solved by tiny Planck scale induced gravitational effects~\cite{Rai:1992xw} or through~\cite{Dvali:1995cc} the symmetry non-restoration at high temperature~\cite{Weinberg:1974hy}. All this said, it is perfectly legitimate to decouple $\Delta_L$, but the naturalness argument is not the right one to use.

Bottom line: in the LR-symmetric seesaw picture that we employ, it is natural, both physically and technically, to work in the limit of vanishing $v_L$ and the $\beta_i$ couplings.

\paragraph*{Case $\mathcal{P}.$} We do not write down explicitly the potential in the case of $\mathcal{P}$. It is enough to say that this case, being more constrained, is  obtained from that of $\mathcal{C}$ by requiring most of the couplings in the potential in~\eqref{CV} to be real. More precisely, a number of phases must vanish and the potential can be obtained from the one in the case $\mathcal{C}$
\begin{eqnarray}\label{P_vs_C}
 &\mathcal{V}_P= \mathcal{V}_C(d_2=d_4=r_4=b_1=b_2=b_3=0) - \nonumber \\
& 2i \alpha_2 s_c \text{Tr} (\tilde{\Phi}\Phi^{\dagger}-\tilde{\Phi}^{\dagger}\Phi)\text{Tr}(\Delta_L\Delta_L^{\dagger})
\end{eqnarray}
We should add that in this case the mass term $\tilde \mu_\Phi$ is automatically real, unlike in the case of $\mathcal{C}$
which required a phase redefinition of $\Phi$.

\subsection{Scalar spectrum.}

Before we go into the gory detail, it is instructive to anticipate the results on physical grounds, at least in the decoupling limit of large $M_{W_R}$ when the spectrum reduces to the SM Higgs boson $h$ with the usual relation for its mass, the heavy triplets $\Delta_{L,R}$ with $m_\Delta^2 \propto \rho v_R^2$ couplings (where $\rho$ stands for the appropriate combination the couplings $\rho_i$)  and of the heavy flavor violating doublet from the bi-doublet with the mass-squared proportional to $\alpha_3 v_R^2$. These essential features get complicated by the possible mixings in the case of accessible scale $M_{W_R}$, but most of them can be understood by symmetry arguments which we present below.

The only relevant relation coming from the first-derivative minimization conditions is for a generic $t_\beta$~\cite{Kiers:2005gh}
\begin{equation}\label{kiers_relation}
t_{2\beta} s_a \simeq -4\frac{\alpha_2}{\alpha_3} s_c\,,
\end{equation}
which holds for both the LR symmetries $\mathcal{P}$ and $\mathcal{C}$. There is an important distinction though.
In the case of $\mathcal{P}$, one has $t_{2\beta} s_a \leq 2 m_b/m_t$~\cite{Senjanovic:2014pva}, which from~\eqref{kiers_relation} implies
\begin{equation}\label{P_restriction}
|2 \alpha_2/\alpha_3 s_c| \leq m_b/m_t\,.
\end{equation}
In the case of $\mathcal{C}$ the parameter $t_{2\beta} s_a$ is unconstrained and no further restriction emerges from~\eqref{kiers_relation}. In both cases, as seen from~\eqref{kiers_relation}, there is no possibility for spontaneous CP violation as opposed to the generic two Higgs doublet situation; the phase $a$ vanishes in the limit of explicit CP conservation ($c=0$). The reason for this is phenomenological, not structural, as we can explain below.

Let us define the following couplings that are useful for the discussion below
\begin{align}\label{l_a}
 \lambda_{\Phi} & \equiv \lambda_1+  s^2_{2\beta}(2\lambda_2c_{d_2+2a}+\lambda_3)+  2s_{2\beta}  \lambda_4c_{d_4+a}\,, \nonumber  \\
  \alpha & \equiv\alpha_1+2\alpha_2s_{2\beta}c_{a+c}+\alpha_3s_{\beta}^2\,, \nonumber \\
  \tilde{\alpha} & \equiv \alpha_2   s_{2\beta} s_a s_c \simeq - 4 \alpha_3 c_{2\beta} (t_{2\beta} s_a)^2\,,
 \end{align}
where $\lambda_{\Phi}$ is  the  quartic coupling of the SM  Higgs if the mixing with $\Delta_R$ fields is neglected, $\alpha$ is the quartic coupling that mixes the SM Higgs with the new Higgs boson in $\Delta_R$ and finally $\tilde{\alpha}$ is the effective quartic responsible of the electroweak corrections to the masses of the $\Delta_L$ multiplets. Notice that $\tilde \alpha$ is negative since $\beta$ is limited due to the perturbativity of Yukawa couplings~\cite{Senjanovic:1979cta}, and is controlled by the physical parameters as we discuss below.

As usual, the next step is to write down the mass matrix through the Hessian of the potential.
It is useful to diagonalize it in two steps: in the first one, we neglect the mixing of the $\Phi$ with $\Delta_R$; in the second one, we consider the whole matrix. Thus we first introduce
\begin{eqnarray}
\phi_{SM} = (c_\beta \phi_1 +s_\beta e^{-ia}\phi_2)=
\left( \begin{array}{cc}
h_{SM}+i G \\
G^- \\
\end{array} \right),
\end{eqnarray}
and
\begin{eqnarray}
\phi_{FV} = (-s_\beta e^{i a} \phi_1 +c_\beta \phi_2)=
\left( \begin{array}{ccc}
H +iA  \\
H^-  \\
\end{array} \right),
\end{eqnarray}
where $\phi_{SM}$ is the SM doublet and ${FV}$ stands for the tree-level flavor violating interactions in which the heavy scalar doublet $\phi_{FV}$  takes part ($\phi_{SM}$ is the doublet with a non-vanishing vev, while $\phi_{FV}$ has a zero vev). In the generic two SM Higgs doublet case these doublets would mix, but here they are eigenstates to a great precision, since $\phi_{FV}$ has to be extremely heavy, on the order of 20 TeV. This allows to ignore the electro-weak contribution of order $v$ to the mass of $\phi_{FV}$.  Moreover, this scalar doublet is basically decoupled, which is why there can be no observable spontaneous CP violation, which as is well known, requires two Higgs doublets with masses at the electro-weak scale~\cite{Lee:1973iz}. Since $m_{\phi_{FV}}^2 \propto \alpha_3 v_R^2$ (see the Tab.~\ref{tab:1}), in order to break CP spontaneously one would need $\alpha_3 \simeq (v/v_R)^2$, clearly in contradiction with the limit on the $\phi_{FV}$ mass. This is made explicit in~\cite{Kiers:2005gh}.

There are possible mixings though with the $\Delta_R$ components (see Appendix~\ref{appendixA}), in particular the mixing between $h_{SM}$ and $\Re e(\delta^0_R)$ is approximatively given by
\begin{equation}
\theta \simeq \frac{\alpha}{2 \rho_1} \frac{v}{v_R}\left[1+\mathcal{O}(\frac{v^2}{v^2_R})\right]\,,
\end{equation}
or more precisely as in appendix~\ref{appendixA}. This mixing is only relevant when the mass of $\delta_R$ in Tab.~\ref{tab:1} is not far from the electroweak scale (small $\rho_1$). Recent limits from electroweak precision tests allow a fairly large $s_{\theta}$ as a function of the mass of the new Higgs~\cite{Falkowski:2015iwa,Godunov:2015nea}, up to $s_\theta \lesssim 0.4$.

In general, the relevant mixing terms among the neutral scalars appearing in Tab.~\ref{tab:1} can be found in Appendix~\ref{appendixA}. Using the constraint in~\eqref{P_restriction}, the expressions in the mass matrix~\eqref{completemassmatrix} get somewhat simplified for the case $\mathcal{P}$, which is reflected in the results given in the Tab.~\ref{tab:1}.

\begin{widetext}
 \begin{center}
 \begin{table}[h]
\begin{tabularx}{\textwidth}[t]{X|X|X}
\arrayrulecolor{black}\hline \hline

\textbf{\textcolor{black}{ Physical scalars }} & \textbf{\textcolor{black}{ Mass$^2$ (case  $\mathcal{C}$) }} & Mass$^2$ (case  $\mathcal{P}$)\\

\hline \hline
$ h \simeq c_{\theta} h_{SM} -s_{\theta} \Re e (\delta_R^0)$ &
\begin{minipage}[t]{\linewidth}%
$4 (\lambda_{\Phi}  -  \frac{\alpha^2}{4\rho_1})v^2$
\end{minipage}&
\begin{minipage}[t]{\linewidth}%
The same with the restrictions in~\eqref{P_vs_C}.
\end{minipage}\\

\arrayrulecolor{black}\hline

 $ \delta_R  \simeq c_{\theta}\Re e (\delta_R^0) +s_{\theta} h_{SM}$ &
\begin{minipage}[t]{\linewidth}%
 $4\rho_1 v_R^2+ \frac{\alpha^2}{\rho_1}v^2$
\end{minipage}&
\begin{minipage}[t]{\linewidth}%
 The same with the restrictions in~\eqref{P_vs_C}.
\end{minipage}\\

\hline

$\phi_{FV}$ (FV heavy doublet) &
\begin{minipage}[t]{\linewidth}%
$\frac{\alpha_3 }{c_{2\beta}} v_R^2 $
\end{minipage}&
\begin{minipage}[t]{\linewidth}%
$\frac{\alpha_3 }{c_{2\beta}} v_R^2 $
\end{minipage}\\

\hline

$\delta_L= \Re e (\delta_L^0)\sim\Im m (\delta_L^0)$ &
\begin{minipage}[t]{\linewidth}%
\qquad$(\rho_3-2\rho_1)v_R^2+4\tilde{\alpha} v^2$
\end{minipage}&
\begin{minipage}[t]{\linewidth}%
\qquad$(\rho_3-2\rho_1)v_R^2$
\end{minipage}\\

\hline

$\delta_L^-$ &
\begin{minipage}[t]{\linewidth}%
$(\rho_3-2\rho_1)v_R^2+(\frac{1}{2}\alpha_3  c_{2 \beta}+4\tilde{\alpha})v^2$
\end{minipage}& \begin{minipage}[t]{\linewidth}%
$(\rho_3-2\rho_1)v_R^2+\frac{1}{2}\alpha_3  c_{2 \beta}v^2$
\end{minipage}\\

\hline

$\delta_L^{--}$ &
\begin{minipage}[t]{\linewidth}%
$(\rho_3-2\rho_1)v_R^2+(\alpha_3  c_{2 \beta}+4\tilde{\alpha})v^2$
\end{minipage}&\begin{minipage}[t]{\linewidth}%
$(\rho_3-2\rho_1)v_R^2+\alpha_3  c_{2 \beta}v^2$
\end{minipage}\\

\arrayrulecolor{black}\hline

$\delta_R^{--}$ &
\begin{minipage}[t]{\linewidth}%
$4\rho_2v_R^2+\alpha_3c_{2 \beta} v^2 $
\end{minipage}&
\begin{minipage}[t]{\linewidth}%
$4\rho_2v_R^2+\alpha_3  c_{2 \beta}v^2$
\end{minipage}\\

\hline
\end{tabularx}
\caption{Physical scalar content of the LRSM and the associated mass spectrum. In the case of $\mathcal{P}$ we discard small terms of $\mathcal{O}(t_{2\beta}s_a)$ for both heavy and light scalars, and in general terms of  $\mathcal{O}(v^2)$ for the heavy flavor changing doublet $\phi_{FV}$. We also ignore small $v/v_R$ corrections which imply that the would-be Goldstone bosons from the light and heavy sectors do not mix. The only phenomenological exception is the mixing $\theta$ which may be non-negligible for light $\delta_R$, in spite of being of order $v/v_R$. Further details on the spectrum and particle mixings are discussed in the Appendix~\ref{appendixA}.}\label{tab:1}
\end{table}
\end{center}
\end{widetext}

We should comment on the results presented above. What is new in Tab.~\ref{tab:1} is the $\beta$ dependence, ignored in the literature by assuming $\tan \beta \simeq 0$. It particularly affects the SM Higgs mass. The $\beta$ dependence enters in the rest of the table mainly through the electro-weak corrections, but it can be important, especially for $\delta_R$ in case it is light, as discussed in subsection~\ref{3c}.

Notice also an interesting fact regarding the sum rule for the masses in the LH triplet, compared to the usual situation of the simple type II seesaw case~\cite{Melfo:2011nx}. The arbitrary sign of the mass splitting is now fixed since $\alpha_3$ must be positive, being responsible for the mass of the heavy $FV$ doublet in the bi-doublet.

\paragraph*{\bf Understanding the spectrum: symmetry arguments.}
\vspace{0.4em}
Let us try to make sense out of the above Tab.~\ref{tab:1} by offering simple symmetry considerations; we believe they ease the reader's pain.

\begin{itemize}
 \item Notice that in the limit $\rho_2 = \rho_4 = c = \alpha_3 = 0$, $\rho_3 = 2 \rho_1$ the masses of the $\Delta$ states vanish, except for $\text{Re}\delta_R^0$. It is easy to understand why this is so, since in this limit the potential exhibits an accidental global $SO(12)$ symmetry which involves 12 real fields in $\Delta_{L,R}$ multiplets. The $SO(12)$ is broken down to  $SO(11)$ through the $v_R$ (assuming $v_L$ =0). Hence 11 Goldstone bosons, the three of them eaten by the heavy gauge fields $Z_R$ and $W_R$. In the $\alpha_3 = 0$ limit the mass of the heavy doublet vanishes too, but it is not due to the $SO(12)$ symmetry arguments. Simply, it is only the $\alpha_3$ that can split the doublets in the bi-doublet since the terms of type $\text{Tr}(\Delta_R^{\dagger}\Delta_R)$ do not affect the $\Phi$ sector. This is what makes the heavy doublet live at the $M_{W_R}$
 scale~\cite{Senjanovic:1978ev} and what cures the usual problem of flavor violation in two-Higgs doublet models~\cite{Senjanovic:1979cta}.

\item In the $\Delta_R$ sector there is a global $SO(6)$ symmetry when $\rho_2 = \alpha_3 = 0$ and once again the $SO(6)$ symmetry is broken down to $SO(5)$ by $v_R \neq0$. There are then 5 Goldstone bosons, three of them are eaten by the gauge fields  $Z_R$ and $W_R^{\pm}$ and the other two  correspond to
$\delta_R^{++}$, which is manifestly massless in that limit. Notice that this is independent of the quartic coupling  $\alpha_2$ which explains why
$\alpha_2$ contribution is absent in the masses of the $\Delta_R$ triplet whereas it appears as a common contribution in all the fields that belongs to $\Delta_L$.

\item It is also instructive to consider the limit $v_R=0$ in which case only $v$ gives mass to the scalars. It gives mass also to both $W$ and $W_R$ (as well the neutral gauge bosons), thus one expects doubling of the Goldstone bosons compared to the SM situation, and it is confirmed by looking at the Tab.~\ref{tab:1} since only the real components of the neutral fields in the bi-doublet pick up masses. There is an interesting exception: $t_\beta=1$ (only a mathematical limit, physically not reachable). In this case only one linear combination of the two $W$'s get massive and thus we expect halving the number of Goldstone bosons in the bi-doublet. An explicit computation confirms it, with $\phi_{FV}$ becoming massive. The limit must be studied apart, it is not smooth.
\end{itemize}

\subsection{The Higgs self-couplings: a window to new physics.}\label{3c}

In the SM the Higgs mass is given in terms of the quartic coupling appearing in the Higgs potential and therefore its determination is a crucial test of the Higgs mechanism. Several studies have been proposed in order to probe the Higgs self-couplings at the LHC and future colliders~\cite{Baglio:2012np,Dolan:2012rv, Gupta:2013zza,Efrati:2014uta,Fuks:2015hna,He:2015spf,Goertz:2013kp,Contino:2016spe}. In particular in~\cite{Gupta:2013zza,He:2015spf} the LHC reach is studied in the context of the scalar singlet extension of the SM, which is effectively the situation encountered in the LRSM for the light $\delta_R$ Higgs scalar.

In Tab.~\ref{tab:3} we give relations between the physical and the original quartic couplings that enter in the scalar masses in Tab.~\ref{tab:1}. We drop the $\sim\theta^2$ corrections in the first two lines, since the forthcoming experiments will not be very sensitive to these interactions.

 \begin{center}
  \begin{table}[h]
\begin{tabular}{ l | c  }
 \hline \hline
 \textbf{\textcolor{black}{ Physical couplings   }} & \textbf{\textcolor{black}{ Quartic couplings }}\\
\hline \hline
$\lambda_{hhhh}$& $ \lambda_{\Phi}/4$  \\
\hline	
$\lambda_{\delta_R\delta_R\delta_R\delta_R}$& $ \rho_1/4$  \\
\hline		
$\lambda_{\delta_R^{++}\delta_R^{++}\delta_R^{--}\delta_R^{--}} $ & $ \rho_1$ \\
\hline
$\lambda_{\delta_L^{+}\delta_L^{-}\delta_L^{+}\delta_L^{-}}-\lambda_{\delta_R^{++}\delta_R^{--}\delta_R^{++}\delta_R^{--}}$ & $\rho_2$\\
\hline
$ \lambda_{\delta_R^{++}\delta_R^{++}\delta_L^{--}\delta_L^{--}}$ & $ \rho_3$ \\
\hline
$ 4\lambda_{\phi_{FV}^{\dagger}\phi_{FV}\delta_L^*\delta_L}-\lambda_{\phi_{FV}^{\dagger}\phi_{FV}\delta_R^{++}\delta_R^{--}}$ & $c_{2\beta}\alpha_3$\\
\hline
\end{tabular}
\caption{Relations among the quartic couplings in the potential and the physical quartic couplings. Small terms of order
$\theta^2$ are ignored in the first two lines.}\label{tab:3}
\end{table}
  \end{center}

The LHC is more sensitive to the triple coupling $\lambda_{hhh}$, since it can be probed in Higgs pair production at the LHC, the reason being that the gluon fusion pair production is the dominant channel  (the order of 30 fb at $\sqrt{s}=14$  TeV~\cite{Baglio:2012np}). The other channels,  such as vector-boson fusion  and associated production with gauge bosons and heavy quarks are generically a factor 10 - 30 smaller. In table~\ref{tab:4} we show the expressions for the relevant trilinear couplings in term of the scalar masses using the relations presented in the Appendix~\ref{appendixA}. A detailed study of the LHC sensitivity to the trilinear coupling~\cite{Baglio:2012np} concluded that the LHC with an integrated luminosity of 3000 fb$^{-1}$ could see the Higgs pair production through the scalar couplings at significant rates. In contrast to the trilinear coupling, the quartic one needs the production of three Higgs bosons in the final state; it is therefore suppressed and probably it cannot be determined at the LHC.

 \begin{center}
 	\begin{table}[h]
 		\begin{tabular}{ l | c  }
 			\hline \hline
 			\textbf{\textcolor{black}{Tri-linear couplings  }} & \textbf{\textcolor{black}{Expression}}\\
 			\hline \hline			
 			$\lambda_{hhh}$& $ \frac{m_h^2}{2 \sqrt{2}} \frac{c^3_{\theta} }{ v}$  \\
 			\hline
 			$\lambda_{\delta_R\delta_R\delta_R}$ & $ \frac{m_{\delta_R}^2}{2 \sqrt{2}} \left(\frac{s^3_{\theta} }{ v}+\frac{c^3_{\theta} }{ v_R}\right) $ \\
 			\hline
 			$\lambda_{h h \delta_R } $ & $\frac{s_{2\theta}c_{\theta}(m_{\delta_R}^2+2m_h^2)}{4\sqrt{2}v} $ \\
 			\hline
 			$\lambda_{h \delta_R \delta_R}$ & $\frac{s_{2\theta}(2m_{\delta_R}^2+m_h^2)}{4\sqrt{2}}(\frac{s_{\theta}}{v}-\frac{c_{\theta}}{v_R})$\\
 			\hline
 		\end{tabular}
 		\caption{Triple scalar couplings in the LR model. We used the relations in the Appendix~\ref{appendixA} to express the tri-linear couplings in terms of the scalar masses. Due to the LHC sensitivity of these couplings, we do not ignore leading $v/v_R$ terms.}\label{tab:4}
 	\end{table}
 \end{center}
 %
 %

Using  the expressions in table~\ref{tab:4} for a quite light $\delta_R$, the trilinear coupling $\lambda_{hhh}$ can be written as~\cite{Gupta:2013zza}
\begin{equation}
\lambda_{hhh} \simeq  \frac{m_h^2}{2 \sqrt{2}}\frac{c^3_{\theta} }{ v}.
\end{equation}
It is instructive to compare it to the SM trilinear coupling, which is  $\lambda^{SM}_{hhh}= \frac{1}{2\sqrt{2}}\frac{m_{h}^2}{v}$ and it gives a deviation with respect the standard model expectation of the form
\begin{equation}\label{deviation}
\Delta \lambda_{hhh} \equiv \frac{\lambda^{SM}_{hhh}-\lambda_{hhh}}{\lambda^{SM}_{hhh}} \simeq 3/2 \theta^2.
\end{equation}
Therefore a  deviation of around 20$\%$ can result for a fairly large $\theta$ (order $\sim 0.4$~\cite{Falkowski:2015iwa}). We shall see in section~\ref{Qver} that this deviation may be much larger once quantum corrections are included\footnote{A complete analysis on the one-loop corrections to the cubic Higgs couplings in the singlet extension of the SM can be found in~\cite{Camargo-Molina:2016moz}.}.
This is encouraging, since at the LHC program with 3000 fb$^{-1}$ of integrated luminosity, the trilinear coupling is expected to be measured with $\pm^{30\%}_{20\%}$ of accuracy~\cite{Goertz:2013kp}.

The prospects for future hadron colliders are even better. For instance, in~\cite{He:2015spf} it is found that a deviation of 13$\%$ can be measured in a 100 TeV collider for 3 ab$^{-1}$ of integrated luminosity, so it is clear that a deviation with respect to the SM values can be found in the present and the next generation of hadron colliders.  Notice that this is complementary to the LNV Higgs decays first considered in~\cite{Gunion:1986im} and phenomenologically investigated within an effective approach in~\cite{Pilaftsis:1991ug}. Recently an in-depth collider study, including displaced vertices, has been provided in~\cite{Maiezza:2015lza} within the LRSM, where this decay is explicitly linkable to the SM deviation of the Higgs boson self-coupling through $\theta$. Furthermore, even if $\delta_R$ is too heavy to be seen at the LHC, for $m_{\delta_R} \gtrsim 2 M_W$~\cite{Nemevsek:2016enw},
the above deviation may be still present for $\delta_R$ below TeV.

{\bf Scalar masses and naturalness.} As discussed above, one can relate directly the scalar masses to the relevant interaction couplings, a general feature of spontaneous broken gauge theories. Trouble occurs for very low scale LRSM though. A $W_R$ in the reach of LHC would require an effective potential beyond tree-level because of the large $\alpha_3$. The issue is analyzed in~\cite{Maiezza:2016bzp} and further discussed in the next sections.

What about the mass scales of various scalar states? First of all, as repeatedly stated the second SM doublet $\phi_{FV}$ is rather heavy, above 20 TeV or so, due to its flavor violating couplings in the quark sector. The left-handed triplet $\Delta_L$ could be light, but for $W_R$ accessible at the LHC it ends up being too heavy to be observed (just as $\delta_R^{++}$), as we discuss in the next section. Ironically, by increasing $M_{W_R}$ the constraints on  $\Delta_L$ and $\delta_R^{++}$ masses go away, allowing them to be light, close to the electro-weak scale. Of course, this become less and less natural as the $W_R$ mass keeps growing. The last remaining state, the RH neutral $\delta_R$ scalar can be as light as one wishes, although again its lightness certainly violates naturalness expectations.

A very light $\delta_R$, with a mass close to the electro-weak scale (and thus decoupled from the RH scale) becomes effectively a SM singlet. This implies a tiny $\rho_1$, so that one looses a direct relation between masses and associated vertices because the latter would be dominated by the Coleman-Weinberg potential~\cite{Coleman:1973jx}. In order to have a predictive theory, one would need the full effective potential, beyond the one in~\cite{Maiezza:2016bzp} that is focused on the leading quantum corrections related to $\alpha_3$. This is explicitly discussed for the relevant trilinear couplings above in section~\ref{Qver}.

{\bf The decoupling limit.} It is worth noticing that in the limit of $\theta\rightarrow 0$ ($m_{\delta_R}\rightarrow \infty$), the expression for the quartic coupling $\lambda_{hhhh}$ in Tab.~\ref{tab:3} does not coincide with the effective quartic coupling appearing in the Higgs mass, with an apparent mismatch with the expected decoupling. The well-known reason is that one must include the reducible diagram $hh\rightarrow hh$ with an intermediate $\delta_R$. Since the relevant trilinear coupling can be expanded from Tab.~\ref{tab:4} as
$\lambda_{hh\delta_R}\simeq 1/{\sqrt{2}}\,\alpha v_R$,
in the decoupling limit one obtains for  the Higgs effective quartic interaction
\begin{equation}\label{lh}
\frac{1}{4}(\lambda_{\Phi}-\frac{\alpha^2}{4\rho_1})h^{4} \equiv \frac{1}{4}\lambda_h h^4.
\end{equation}
This is precisely the same effective quartic entering in the Higgs mass $m_{h}^2=4\lambda_{h}v^2$ in Tab.~\ref{tab:1}, where $\lambda_\Phi$ and $\alpha$ are defined in~\eqref{l_a}.

\section{The scalar potential at work}\label{IV}

In this section we examine the behavior of the potential under the running in the whole parametric space of the model for three different LR scales: LHC reach, next collider reach (i.e.$\sim 20$ TeV) and very high energy ($10^{9}-10^{11}$ GeV). The complete renormalization group equations (RGE) for the quartics were first provided by~\cite{Rothstein:1990qx}  and recently revisited in~\cite{Chakrabortty:2016wkl}, where some constraints on the parametric space are derived.

\begin{figure*}[t]
	\centerline{%
\hspace{-0.9em}
		\includegraphics[width=.66\columnwidth]{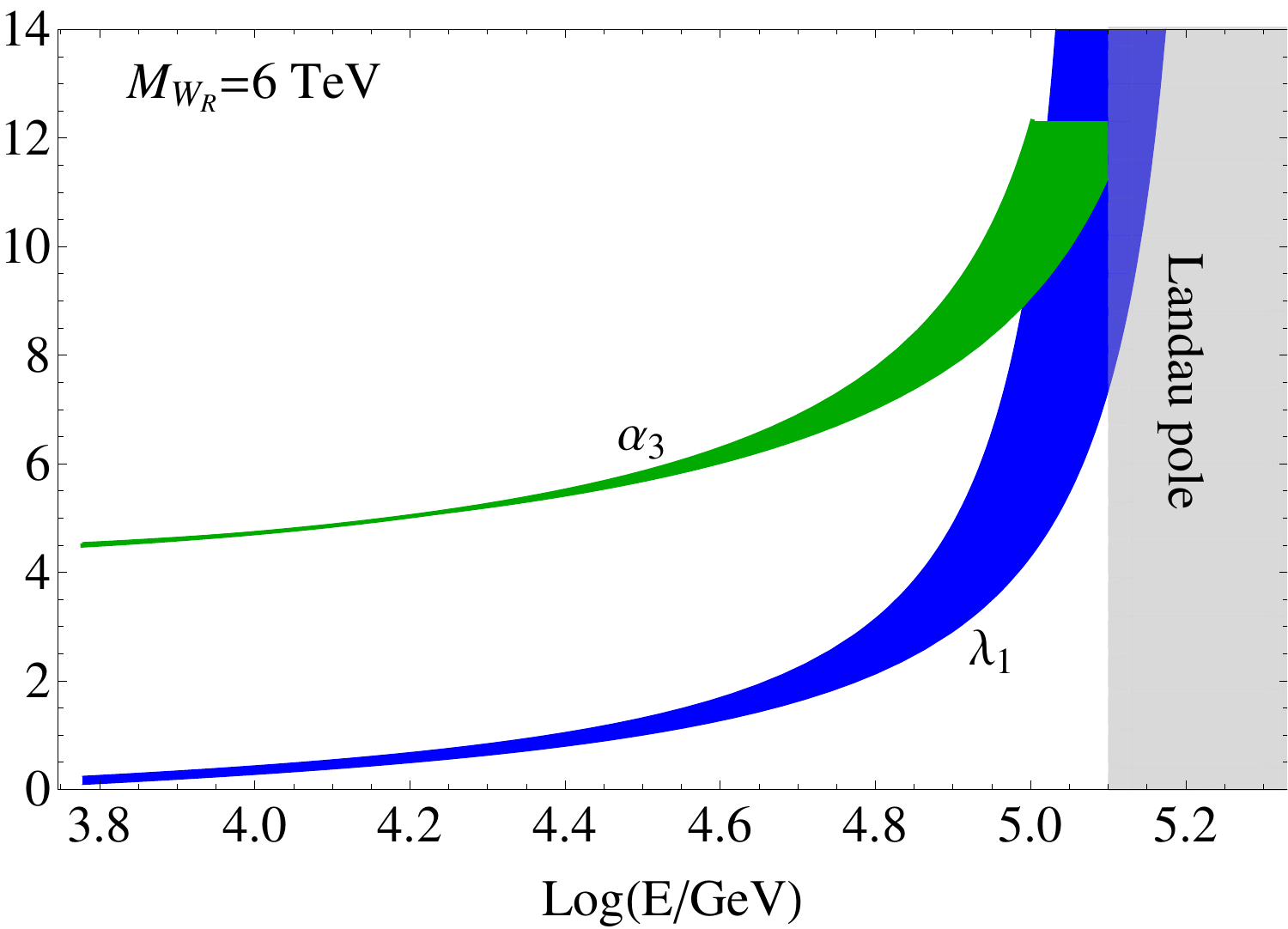}~~~~
\hspace{-2.06em}
		\includegraphics[width=.73\columnwidth]{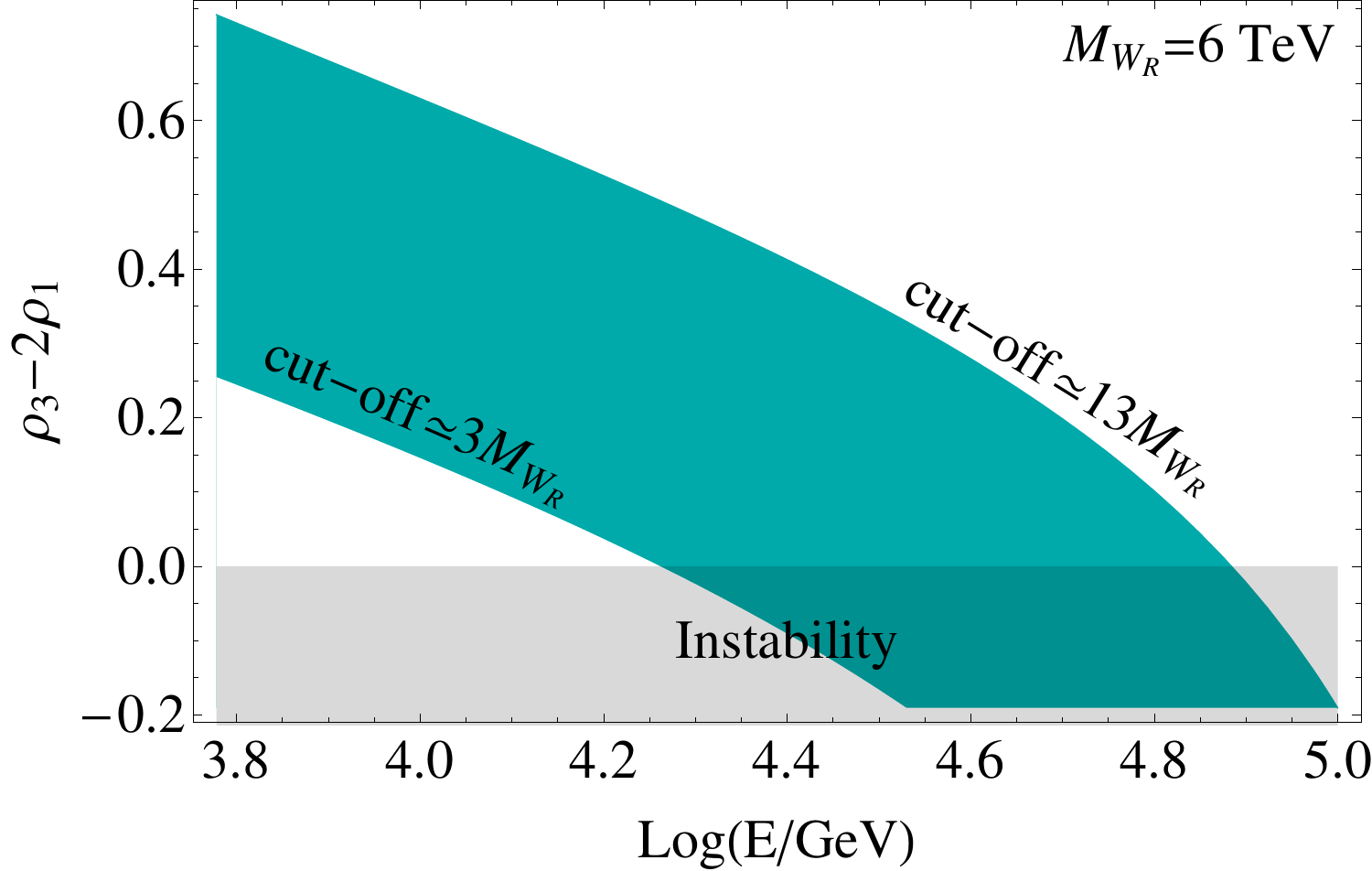}%
\hspace{-0.53em}
		 \includegraphics[width=.73\columnwidth]{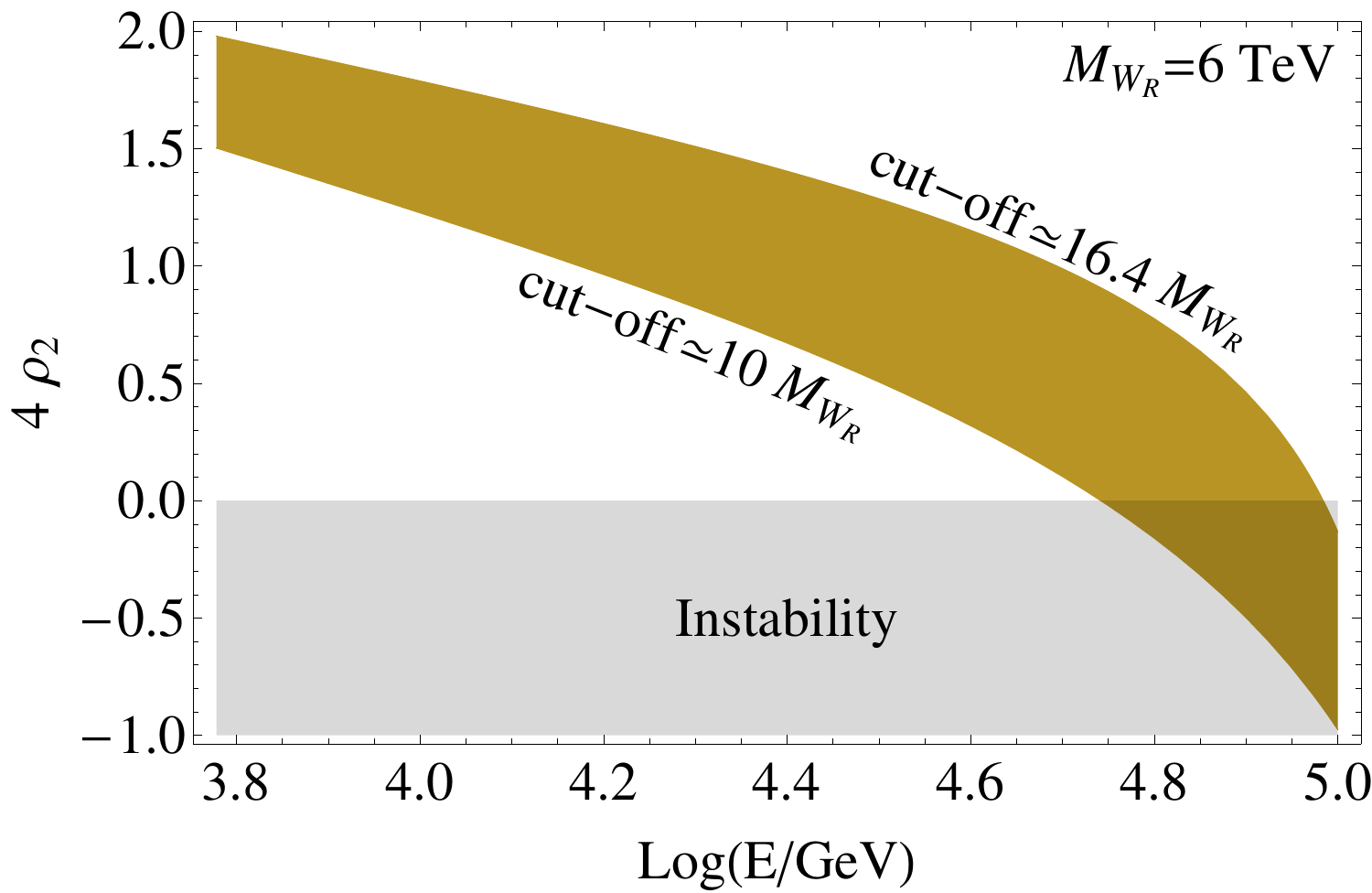}%
	}%
	\vspace*{-2ex}
	\caption{Left. Running of $\lambda_1,\alpha_3$ (the other $\lambda$ and $\alpha$ couplings exhibit a similar behavior), they become non-perturbative around $10^5$ GeV. Center. Running of $\rho_3-2\rho_1$ which provides the leading masses for the $\Delta_L$ multiplets. The values for the cut-off are read off from the point where  $\rho_3-2\rho_1$ goes to zero. Right. The same for $4 \rho_2$ which provides the leading mass term for $\delta_R^{++}$.  In all plots the bands denote the dependence on the random initial choices consistent with the mass spectrum.}
	\label{LHCfig}
\end{figure*}

For low LR scale one has to deal with a large $\alpha_3$ required by the heaviness of the doublet $\phi_{FV}$.
The the finite one-loop contributions to the generation of the quartics at $v_R$ due to the large $\alpha_3$ coupling were taken systematically into
account in~\cite{Maiezza:2016bzp}. Here we consider the divergent loop contributions and the running of the couplings by choosing randomly the initial quartics (consistently with the expression for the masses). Moreover, we allow the possibility of $t_\beta\neq0$ since it enters directly in the RGE's and more important, it changes drastically the matching conditions of the starting quartics with the scalar masses in Tab.~\ref{tab:1}. It is not justified to
set the initial values of $\lambda_{2,3,4}$  to zero as in the present literature since these couplings contribute to the Higgs mass $m_h$ as clear from Tab.~\ref{tab:1} and~\eqref{l_a} and furthermore they are not self-renormalizable.

We extend the analysis for the LR scale at next collider generation, where the LRSM is less constrained, showing that the theory becomes completely natural and remains perturbative all the way to high scales. Finally, we consider the case of very high energy RH scale, relevant for the two step symmetry breaking of the $SO(10)$ GUT. It is crucial to make sure that the theory remains perturbative in the energy regime between the intermediate LR breaking scale and the scale of grand unification. As we shall see, this can be satisfied as long as the scalar states tend to live below the LR scale.

\begin{figure*}[!]
\centerline{%
\includegraphics[width=.97\columnwidth]{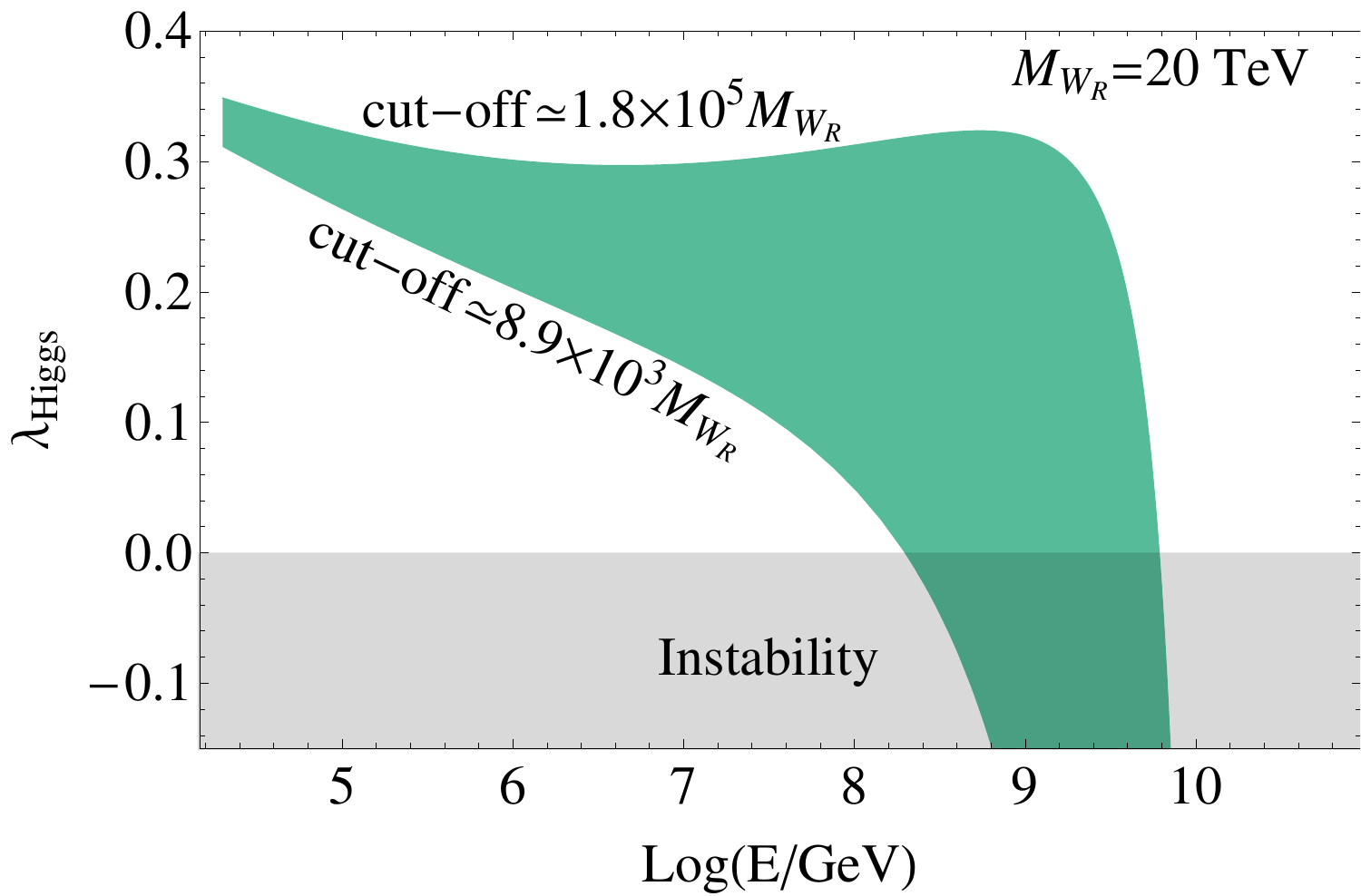}~~~~
\includegraphics[width=.94\columnwidth]{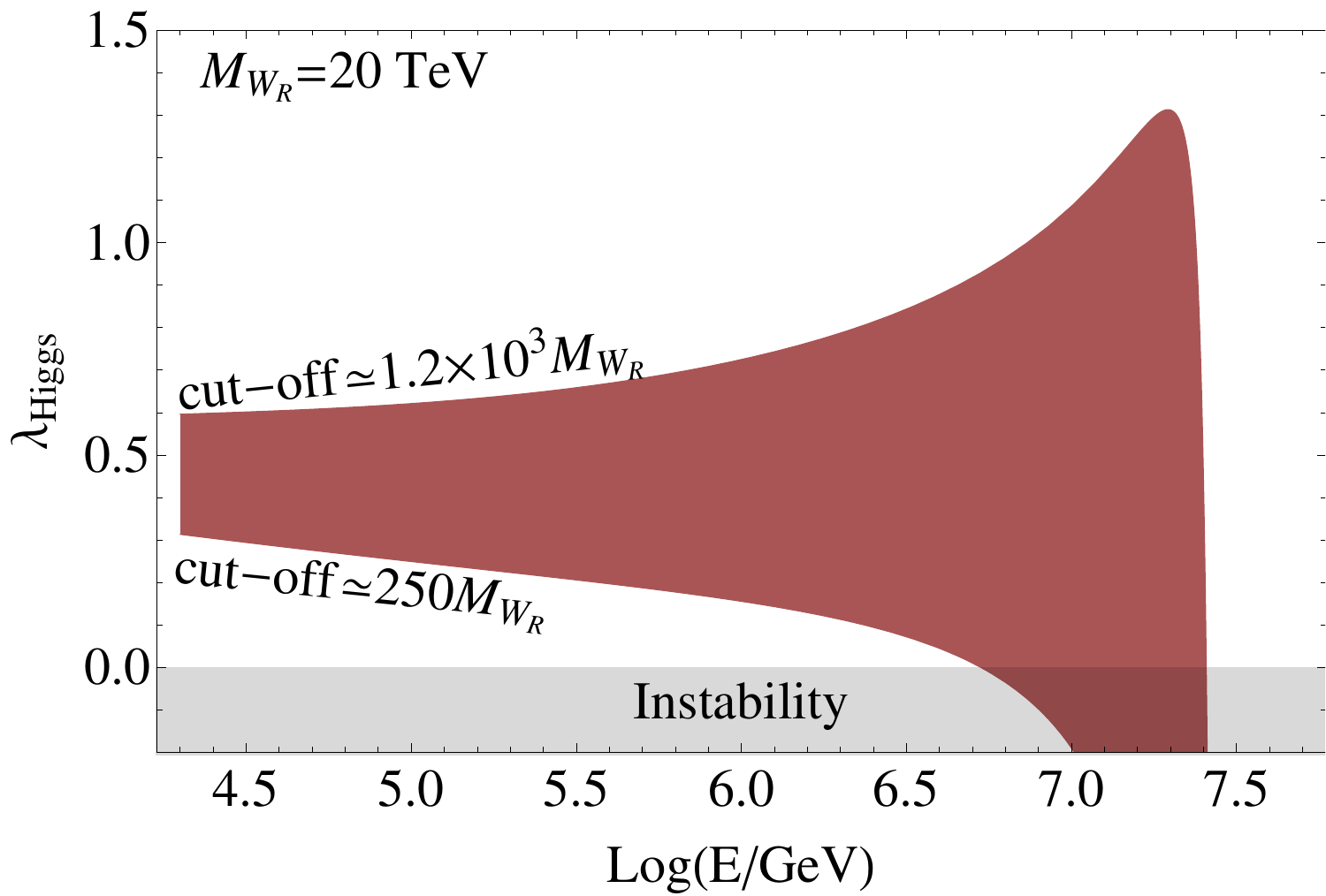}%
}%
\vspace*{-2ex}
\caption{Left. Running of $\lambda_{Higgs}\equiv 4 \lambda_h$ defined in~\eqref{lh} for $t_\beta$=0.
Right. The same for $t_\beta$=0.3 giving a lower cut-off.
The cut-offs are defined in the same manner as in the Fig.~\ref{LHCfig}}\label{nextcolliderfig}
\end{figure*}

\subsection{Left-Right symmetry at LHC}

Let us start our discussion on RGE's in the phenomenologically most relevant case of low RH scale. As already remarked above, in this case the scalar
potential is strongly affected by the large $\alpha_3$ and its induced quantum effects. The evaluation of the self-induced $\alpha_3$ at one-loop and
at $v_R$ scale yields~\cite{Maiezza:2016bzp} $\alpha_3^{(1)}/\alpha_3 = 3 \alpha_3/(8\pi^2)$,
which means a perturbativity of $\sim10\%$ for $M_{W_R}\simeq6$ TeV (the value for which the perturbativity issue is maximally alleviated, while $W_R$ is still detectable at LHC~\cite{Ferrari:2000sp}). Therefore we focus exactly on this portion of parameter space of the model, which means $\alpha_3\lesssim 5$~\cite{Bertolini:2014sua, Maiezza:2016bzp}. Taking the lower limit as an input and choosing the other quartics randomly within the range\footnote{The analysis is done even by choosing randomly negative values for those quartics not responsible for the leading mass terms of the scalars. No significative differences emerge.} (0,0.1) but consistently with the spectrum, several couplings become non-perturbative above $10^5$ GeV.

The running of $\lambda_1,\alpha_3$ is shown in Fig~\ref{LHCfig}(Left). The result depends on the random choice of the initial quartics while being quite insensitive to $t_\beta$. Increasing the range to be (0,1), the situation worsens and the Landau pole of the theory gets too low.
The cutoff from Fig.~\ref{LHCfig}(Left) is lower than the one shown in~\cite{Chakrabortty:2016wkl}, due to the larger initial $\alpha_3$.
In the running, the threshold effects are taken into account, the light scalars start to run below $v_R$ at their own mass values.

Other important results of the RGE's of the scalar sector with the RH scale at LHC is represented by in Figs.~\ref{LHCfig}(Center,Right). The combination $\rho_3-2\rho_1$  provides the leading mass term of the $\Delta_L$ components (see Tab.~\ref{tab:1}). The parameter $\rho_3-2\rho_1$ can become negative as in Fig.~\ref{LHCfig}(Center), destabilizing the potential below the limits from perturbativity in Fig.~\ref{LHCfig}(Left). In order to get the cut-off (defined as the point where this parameter vanishes) as far as possible above $M_{W_R}$, one has to choose those configuration where the initial $\rho_3-2\rho_1$ is large enough, without worsening significantly the perturbative limit (Landau pole).

{\bf Theoretical limits on the masses of the triplet components.}
In terms of the masses of $\Delta_L$ triplet, for the chosen value $M_{W_R} = 6$ TeV, this arguments reads from the
Fig.~\ref{LHCfig}(Center) as
\begin{equation}\label{mDL}
\text{cutoff}\gtrsim 10 \,M_{W_R} \,\, \Rightarrow \,\,  m_{\delta_L,\delta_L^+,\delta_L^{++}}\gtrsim 9\,\text{TeV}.
\end{equation}
This is not the actual limit on the masses of $\Delta_L$. It only applies to the $W_R$ accessible at LHC, while for a $W_R$ mass above roughly 20 TeV, it goes away completely, as we discuss in the next subsection. Physically, it says that if the $W_R$ were to be discovered at the LHC, $\Delta_L$ should not be seen.

Exactly the same discussion applies for Fig.~\ref{LHCfig}(Right) that shows the running of the quartic $\rho_2$, related to the leading mass term of $\delta_R^{++}$. One has to choose the initial $\rho_2\gtrsim 0.35$ consistently with the cut-off in Fig.~\ref{LHCfig}(Right) and without spoiling significantly the Landau pole in Fig.~\ref{LHCfig}(Left), thus
\begin{equation}\label{mDR}
m_{\delta_R^{++}}\gtrsim 12\,\text{TeV}.
\end{equation}
These LHC constraints are stronger than the phenomenological ones from the oblique parameters in~\cite{Maiezza:2016bzp}, and larger than the benchmark values considered in~\cite{Bambhaniya:2013wza}.

We believe that a cut-off as in~\eqref{mDL} is the smallest value for living safe, just enough to suppress non-renormalizable operators from a new physics scale, at least in those configurations with $\text{cutoff}\gtrsim 10 \,M_{W_R}$. The model requires though a UV completion already in the reach of the next collider generation, which can be seen as a challenge. Still, the conclusion is that the entire scalar content of the LRSM has to be heavy, except for $\delta_R$ that remains unconstrained. This is crucial in relation with the discussions encountered in section~\ref{3c}. We should stress that by lowering the $W_R$ mass, the cut-off goes down, below the order of magnitude limit we used as a definition of a sensible renormalizable theory. This implies $M_{W_R} \gtrsim 6 \,\text{TeV}$, whereas, as remarked before, the LHC reach requires~\cite{Ferrari:2000sp} $M_{W_R} \lesssim 6 \,\text{TeV}$ - at the LHC the theory lives at the edge.

A final comment is in order. What is exhibited in Fig.~\ref{LHCfig}(Center, Right) represents proper instabilities (not meta-stabilities), since the estimated decay time from~\cite{Lee:1985uv} is very short with respect to the age of the universe. The same holds for the instabilities discussed below.

\subsection{Left-Right symmetry at next hadronic collider}

The proper machine for the LRSM would be a 100 TeV collider in any case, since the $FV$ scalar doublet is far away from the LHC reach. Therefore we choose to focus in this section on the LRSM with $M_{W_R}=20$ TeV, consistent with next generation colliders. This choice, besides eliminating any tension in the parametric space of the model, represents a scale for which the LRSM offers an insight on the strong CP problem. Namely, the restoration of parity makes $\bar{\theta}$ computable~\cite{Mohapatra:1978fy} leading to $M_{W_R}\geq20$ TeV~\cite{Maiezza:2014ala}. This also fits well with the potentially strong limit due to $\epsilon'$~\cite{Bertolini:2012pu}.

The general setup of the RGE analysis is the same of the one discussed in the previous subsection, except  that now $\alpha_3$ can be fairly small. From the $FV$ constraints one has $\alpha_3\gtrsim 0.38$~\cite{Bertolini:2014sua,Maiezza:2016bzp}, being the lower value our input parameter.

The most stringent limits are obtained by the running of $\lambda_{Higgs}\equiv 4 \lambda_h$ defined in~\eqref{lh}, and they depend on $t_\beta$. In the left panel of Fig.~\ref{nextcolliderfig} we choose $t_\beta =0$, leading to a destabilization of the potential around $10^9$ GeV. The result is seen to depend on the random choices of the initial values, and we conservatively quote the worst configuration.

A non-vanishing $t_\beta$ enters directly in the RGE's through the Yukawa couplings and the cut-off gets lowered. For $t_\beta =0.3$, chosen for the sake of illustration\footnote{Larger values imply a less perturbative interaction of the $FV$ scalars with the quarks~\cite{Senjanovic:1979cta}.}, the potential is destabilized around $10^7$ GeV, as shown in the right side of Fig~\ref{nextcolliderfig}.

As a result, we believe that it is not well motivated to focus on versions of the theory in which parity is broken at very high energy, while the gauge symmetry $\mathcal{G}_{LR}$ is preserved up to 10-100 TeV - at least, not by appealing to grand unification. The quartic couplings become non-perturbative well below the GUT scale and this holds even for the truncated potential~\cite{Chang:1983fu,Dev:2016dja} consistent with the high scale parity breaking picture.

In short, a RH scale in the range 10-100 TeV leads to a well-defined perturbative model, with a high scale cutoff. Moreover, the theoretical limits on the masses of $\Delta_L$ states and $\delta_R^{++}$ are now gone away and one is left only with the experimental bounds on the order of a few hundred GeV.

\subsection{High scale Left-Right symmetry and $SO(10)$ GUT}

\begin{figure*}[t]
\centerline{%
\includegraphics[width=.97\columnwidth]{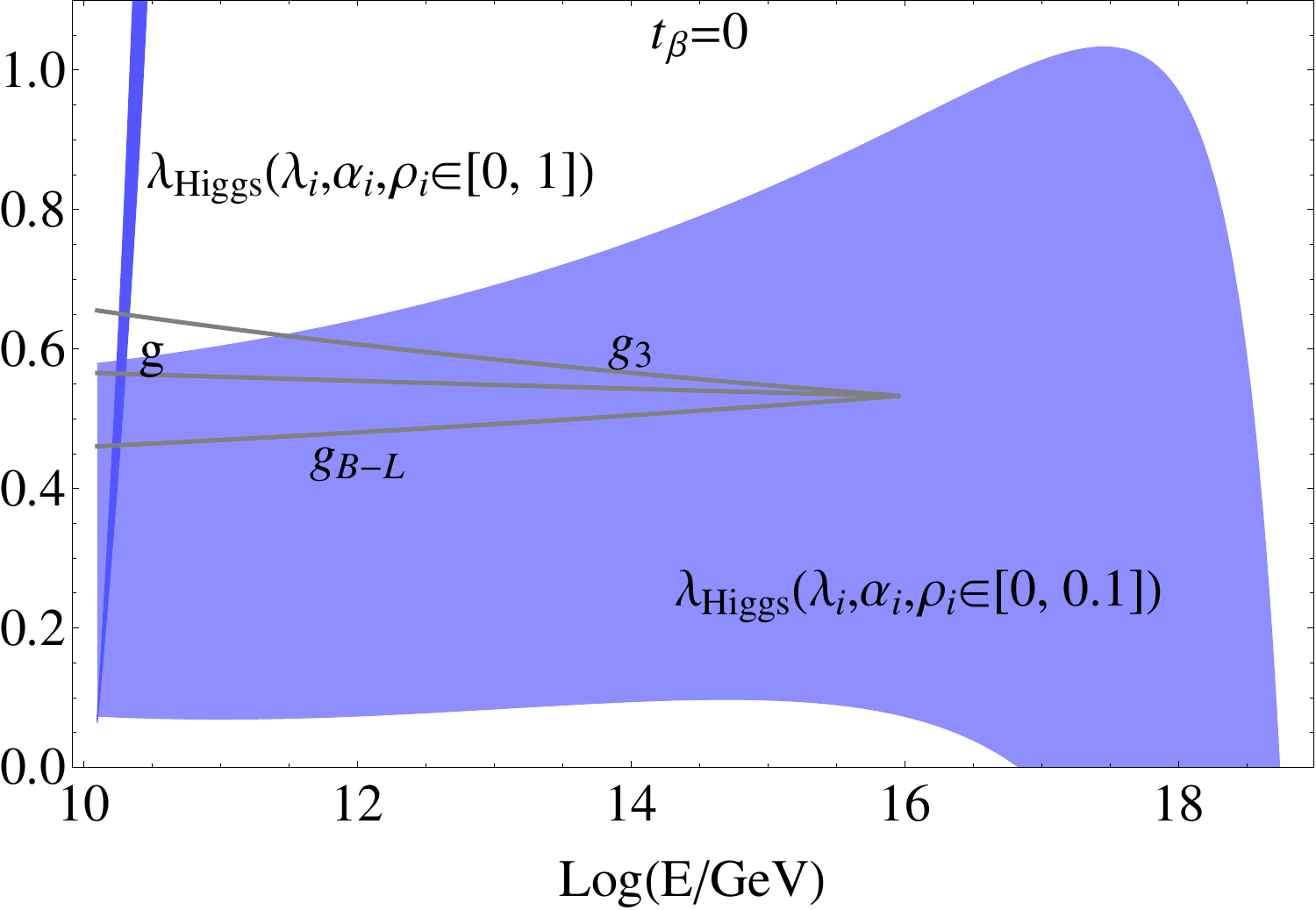}~~~~
\includegraphics[width= .95\columnwidth]{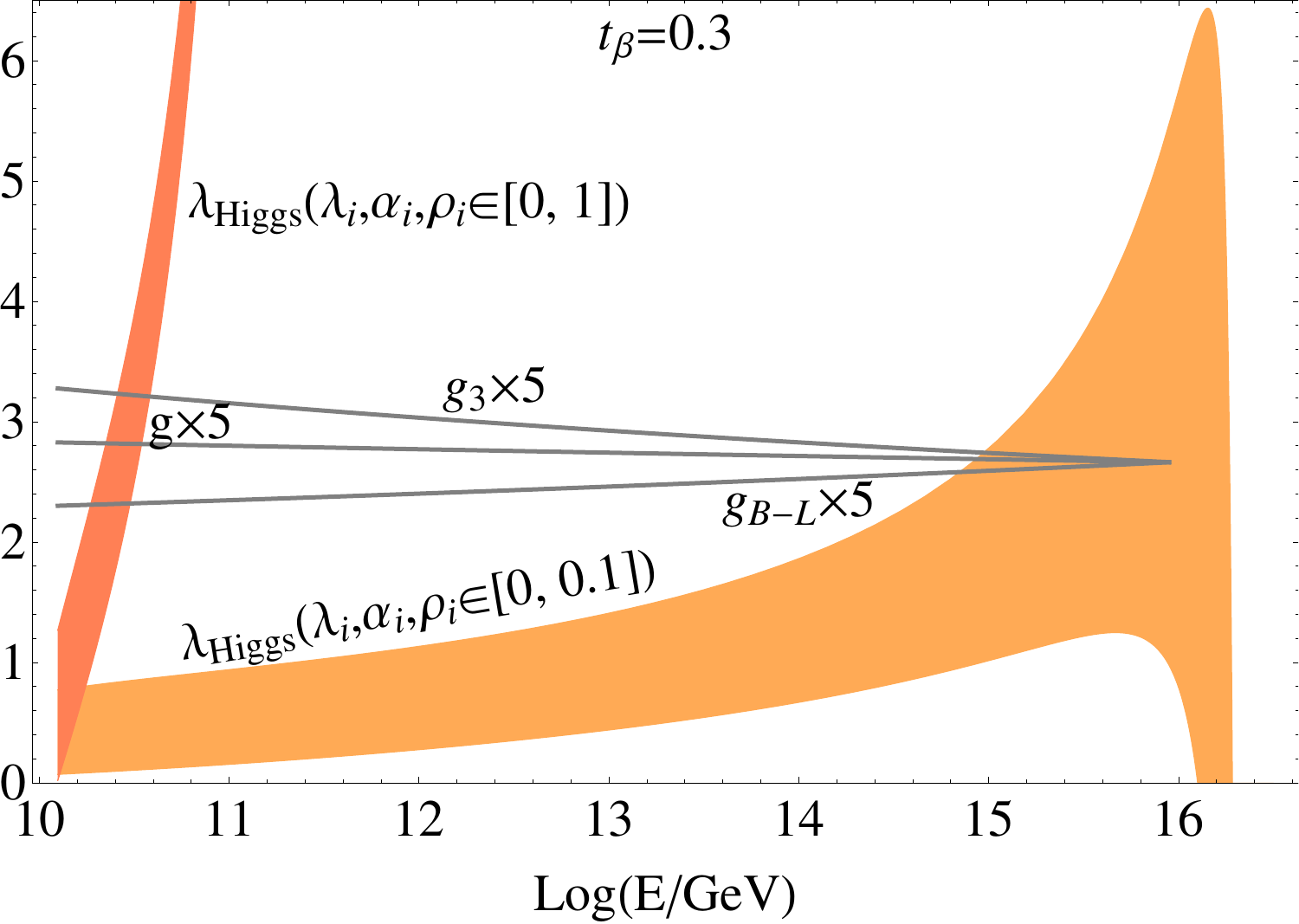}%
}%
\vspace*{-2ex}
\caption{Left. Running of $\lambda_{Higgs}\equiv 4 \lambda_h$ defined in~\eqref{lh} for $t_\beta$=0.
Right. The same for $t_\beta$=0.3 which shows a lower cut-off and $\lambda_{Higgs}$ can become slightly large at GUT scale. The cut-offs are defined as in the previous figures.}\label{GUTfig}
\end{figure*}

The LRSM can be naturally embedded in the $SO(10)$ GUT with the generalized charge conjugation $\mathcal{C}$ a discrete $SO(10)$  gauge symmetry.
With the minimal fine-tuning hypothesis, the LR and GUT scales are predicted to be $\sim10^{10}$ GeV and $\sim10^{16}$ GeV respectively~\cite{delAguila:1980qag}. A question arises naturally: are there any conditions on the scalar potential needed to ensure the consistence of this picture? After all, the quartics of the potential have to remain perturbative up to the scale of grand unification.

In the Fig.~\ref{GUTfig} we illustrate once again the cases of $t_\beta=0; 0.3$ for two different ranges of the quartics. For the sake of completeness we plot also the gauge couplings as a benchmark. By varying randomly the initial quartics, one sees that the two step $SO(10)$ symmetry breaking can be preserved with $|\lambda_i,\alpha_i,\rho_i|<0.1$, albeit non-trivially. The case of non-null $t_\beta$ is slightly disfavored, as clear from the right side of Fig.~\ref{GUTfig}. In fact, although the cut-off is still around GUT scale, $\lambda_{Higgs}$ can become fairly large below the destabilization point of the potential.

In any case, keeping the quartics of order of few percent is sufficient to preserve the standard $SO(10)$ picture. This implies that the scalar masses tend to be lower than $v_R$, in reasonable accord with the extended survival principle (equivalent to minimal fine-tuning) needed in order to make predictions on the mass scales in grand unification~\cite{survival}. In short, all is well with the naive picture, as long as the scalars live somewhat below the corresponding symmetry breaking scale.

{\bf Higher order effects.} Before closing this section, a discussion is needed regarding higher order effects. The one-loop RGE's for the LRSM show fairly large coefficients in the pure quartics part~\cite{Rothstein:1990qx} due to the rich scalar field content. One has to wonder whether at higher orders even larger coefficients appear, breaking down the perturbative expansion and spoiling the one-loop results. A complete two-loops analysis is beyond the scope of this work. Still, it is important to check the impact on the running from this main part of $\beta_{2-loop}$ related to the quartics only. In the Appendix~\ref{appendixB}, as generic example, we show  the $\beta$-function for $\lambda_1$ at the two-loop order.

As can be seen from~\eqref{1_2_match}, no relevant impact on the running is expected in the cases of very high energy RH scale and next collider reach, since there the quartics can be fairly small (we verify this by explicit calculation). In the case of LR symmetry at LHC, $\alpha_3$ is large and so most of the other couplings grow quickly during the running. A direct evaluation shows that the two-loop correction reduces a bit the already low destabilization point. However, the Landau pole appears still slightly above the cut-off shown in Figs.~\ref{LHCfig}(Center, Right), which in turn is not drastically modified. In conclusion, the results presented in this section are quite stable.

\section{Trilinear vertex corrections}\label{Qver}

\begin{figure*}[t]
\centerline{%
\includegraphics[width=.495\columnwidth]{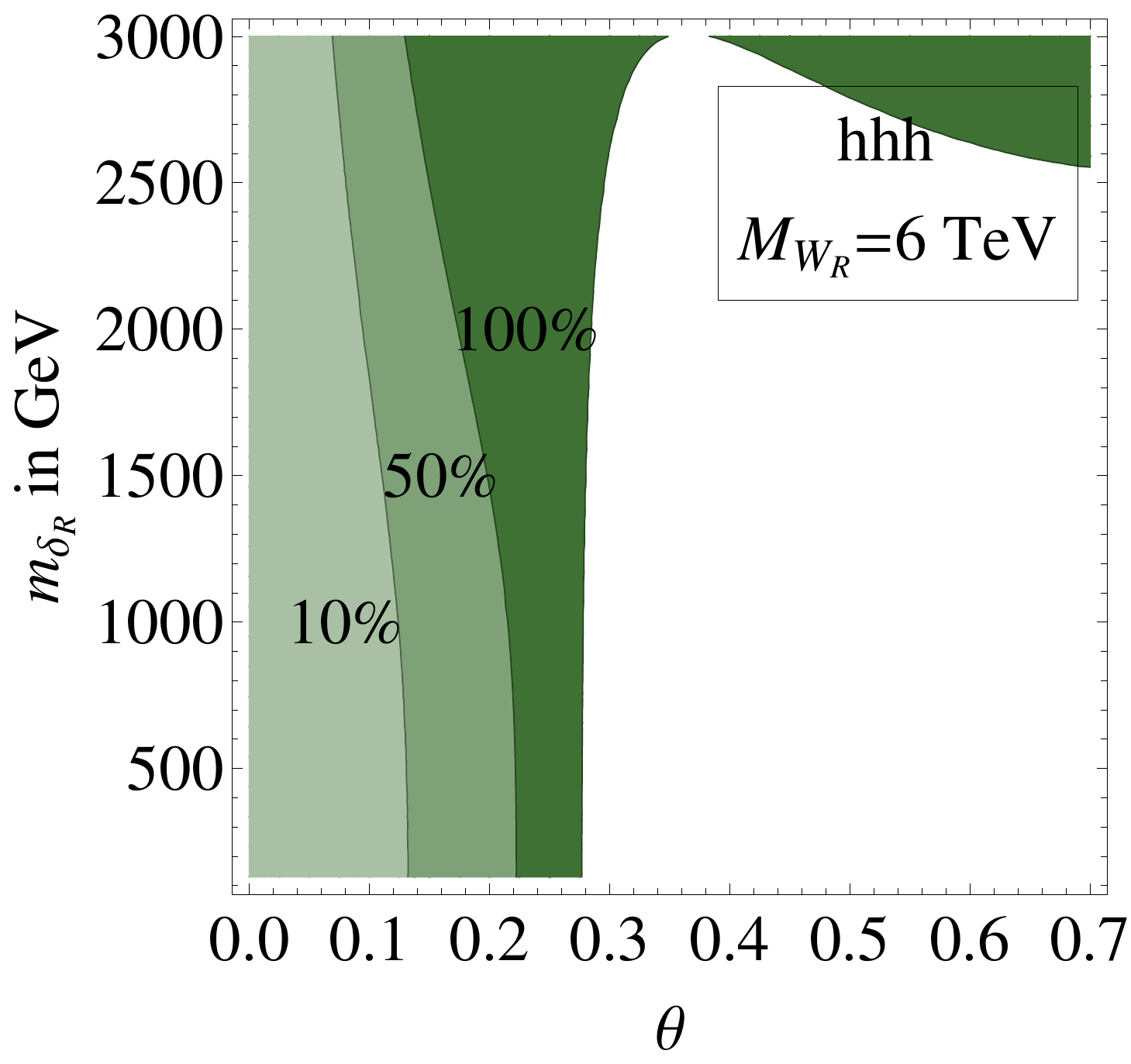}~~~~
\includegraphics[width= .495\columnwidth]{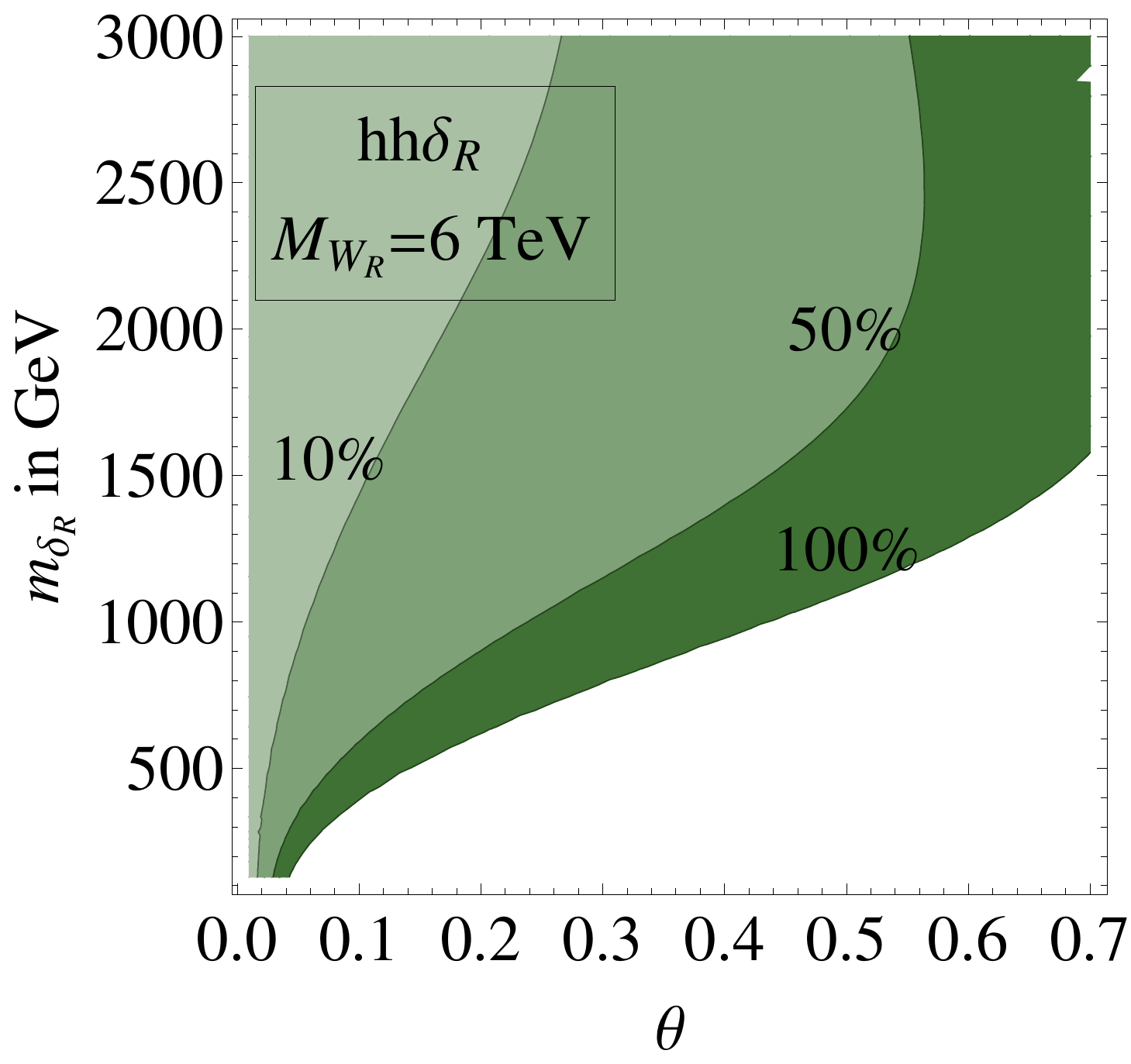}%
\includegraphics[width= .495\columnwidth]{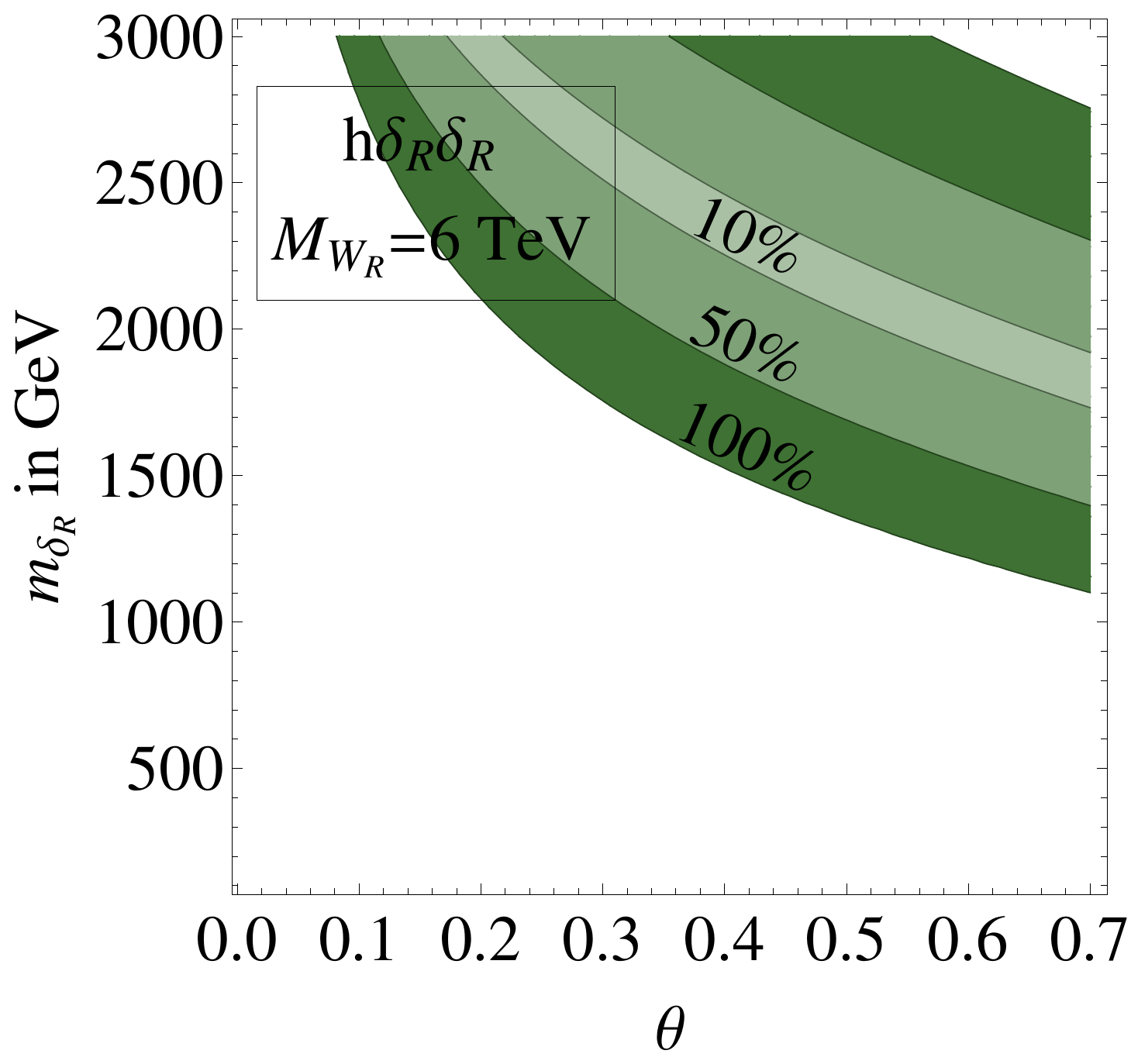}%
\includegraphics[width= .495\columnwidth]{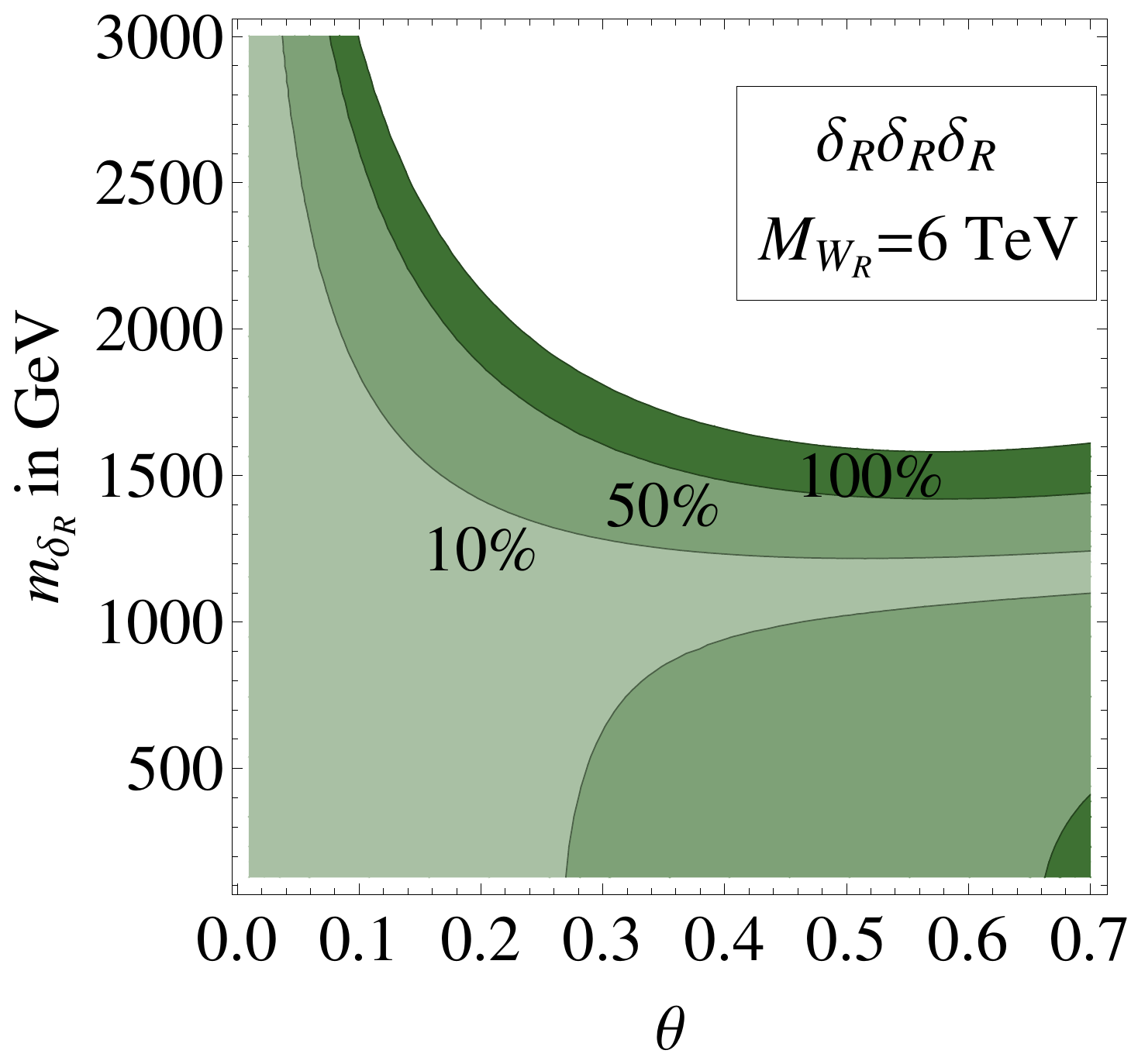}%
}
\centerline{%
\includegraphics[width=.495\columnwidth]{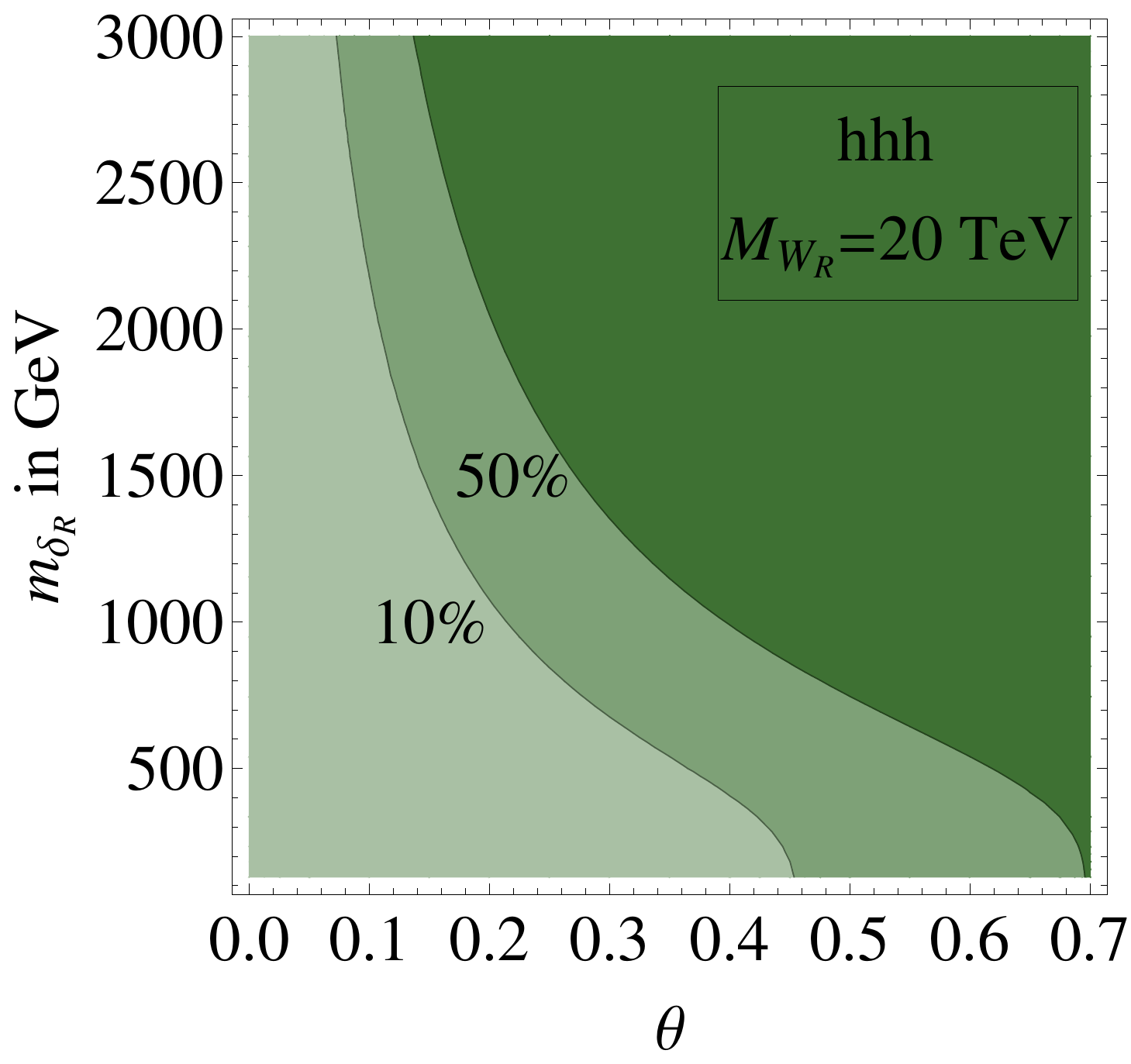}~~~~
\includegraphics[width= .495\columnwidth]{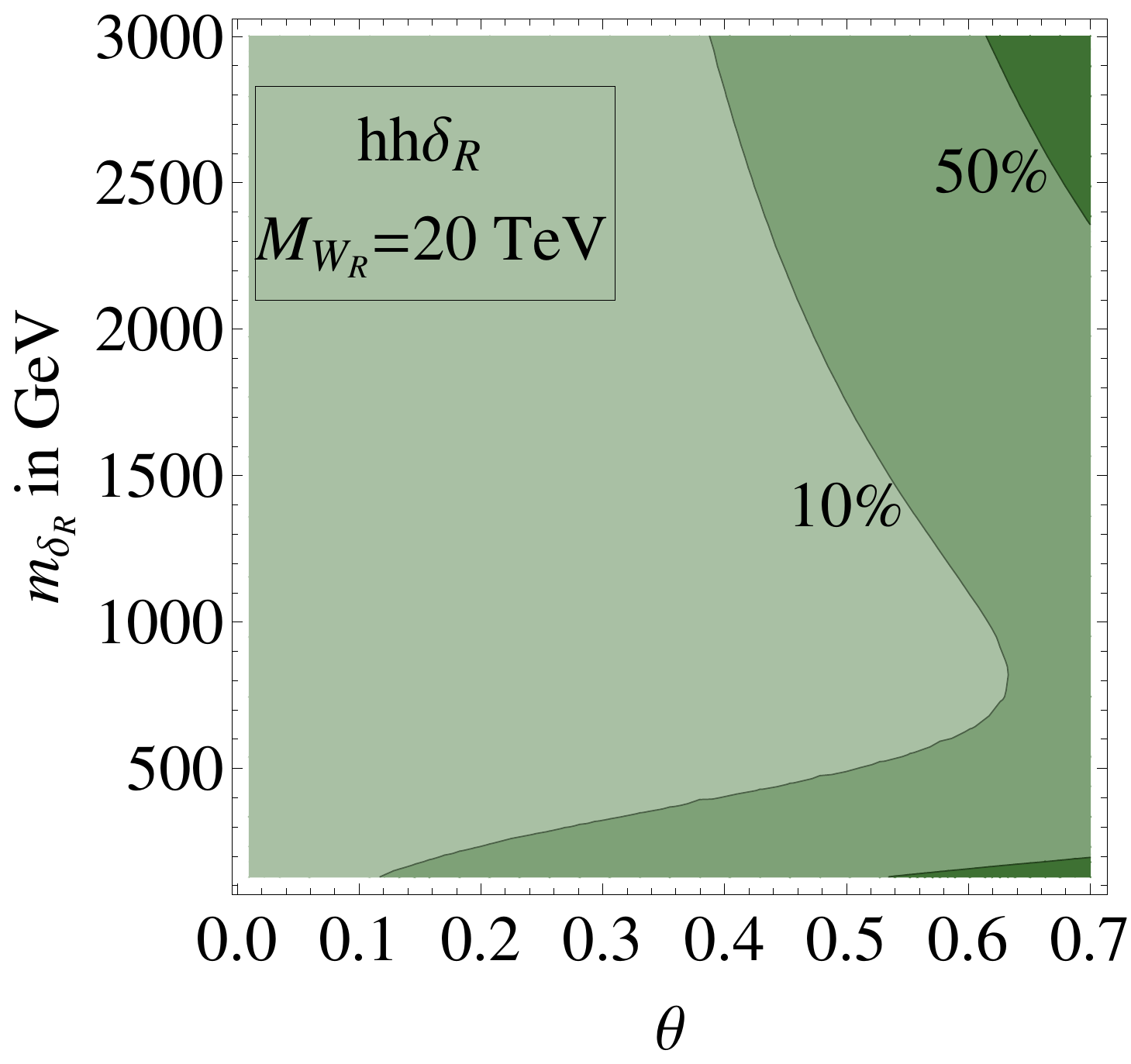}%
\includegraphics[width= .495\columnwidth]{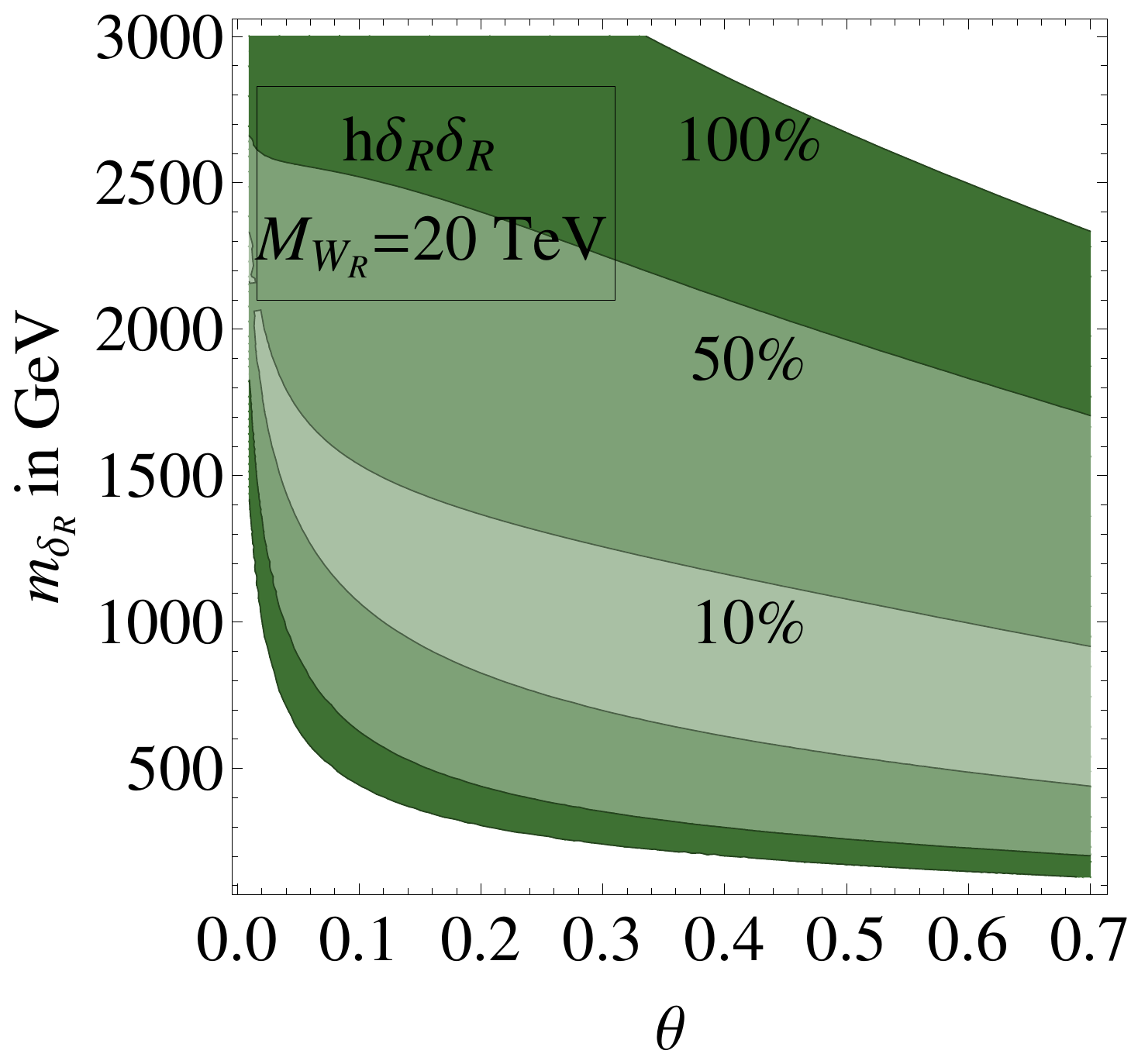}%
\includegraphics[width= .495\columnwidth]{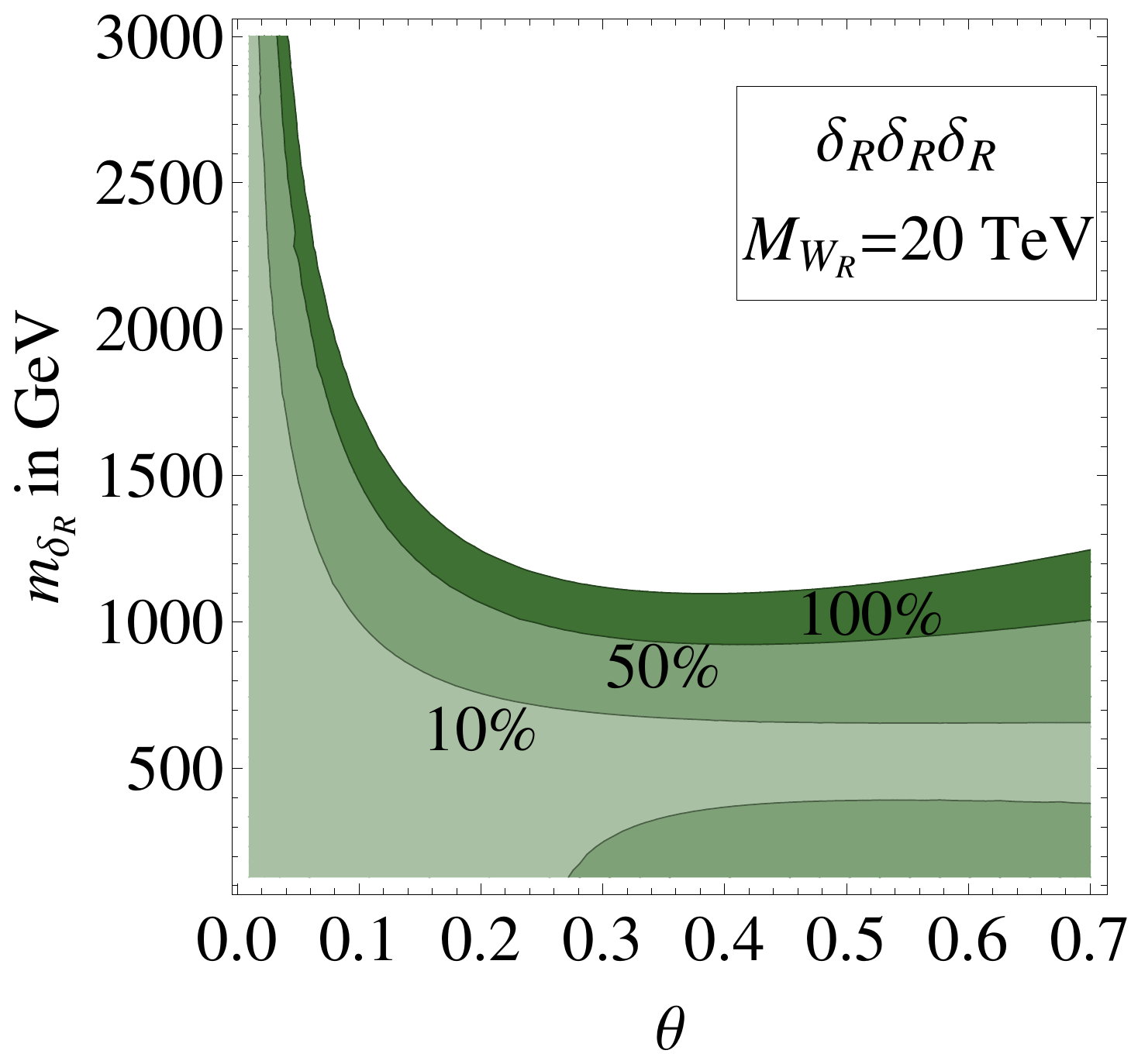}%
}
\vspace*{-2ex}
\caption{Plots for the quantities shown in~\eqref{loop_deviation} (in $\%$) for $M_{W_R}=6$ TeV (Top) and $M_{W_R}=20$ TeV (Bottom). For the sake of clearness the plots run up to $\theta\simeq 0.7$, although some regions are ruled out phenomenologically~\cite{Falkowski:2015iwa}.}\label{quantum_cor_fig}
\end{figure*}

Here we discuss the one-loop renormalization vertex for the cubic couplings; similar results hold also for the quartic couplings.
Of particular importance is the limit of $m_{\delta_R}\ll M_{W_R}$, since  a phenomenologically appreciable impact on the Higgs physics requires a light $\delta_R$, partially decoupled from RH scale. This, in turn, implies domination of the quantum corrections for the trilinear and quartic  couplings involving $\delta_R$ in the effective potential. Moreover, a $W_R$ in the reach of LHC requires a large $\alpha_3$ and therefore its related loop effects may be the dominant ones. In this case the leading quantum correction can be read off from the effective potential in~\cite{Maiezza:2016bzp}, where in particular one sees the trilinear $\delta_R^3$ coupling (re-scaling $\delta_R\rightarrow \delta_R/\sqrt{2}$ in usual normalization) $\simeq \left(\sqrt{2} \rho_1  +2\sqrt{2}\, \alpha_3^2/(48 \pi^2)\right)v_R$.

Clearly, for $\delta_R$ sufficiently light (small $\rho_1$), the loop effect becomes dominant. One should not confuse this with the perturbativity issue in the LRSM  discussed  in~\cite{Maiezza:2016bzp}; simply the perturbation theory starts at the one-loop level when the tree level is made artificially small, as known from the classic work of Coleman and Weinberg~\cite{Coleman:1973jx}.

We consider here the quantum corrections to the tree-level exact expressions in the Table~\ref{tab:4} and the Appendix~\ref{appendixA} by including the whole scalar spectrum. The latter is  especially important in the case of the RH scale in the LHC reach, where one has to consider even the constraints in~\eqref{mDL} and~\eqref{mDR}.
The complete expressions of the effective trilinear couplings are too long to be reported here, thus we show the leading corrections to the expressions in Table~\ref{tab:4} in the limit $\theta,\rho_1\rightarrow 0$:
\begin{align}
&\lambda_{hhh}^{approx}= \lambda_{hhh}+\nonumber \\ & \frac{1}{\pi ^2}\left[\frac{v^3}{v_R^2} \left(\frac{\sqrt{2} \lambda _{\Phi}^3}{3 \alpha _3}+\frac{\alpha _3^3}{96 \sqrt{2} \rho _2}+\frac{3 \alpha _3^3}{64 \sqrt{2} \rho _3}\right)+\frac{9 \lambda _{\Phi}^2 v}{8 \sqrt{2}}\right], \label{hhh_aprox}\\
&\lambda_{hh\delta_{R}}^{approx}= \lambda_{hh\delta_{R}}+\frac{v^2 \left(9 \alpha _3^2+32 \text{$\lambda_{\Phi} $}^2\right)}{32 \sqrt{2} \pi ^2 v_R}, \label{hhd_aprox}\\
&\lambda_{h\delta_{R}\delta_{R}}^{approx}=\lambda_{h\delta_{R}\delta_{R}}+\frac{\alpha _3 v \left(8 \left(\lambda _{\Phi}+\rho _2\right)+3 \rho _3\right)}{16 \sqrt{2} \pi ^2},\label{hdd_aprox}\\
&\lambda_{\delta_{R}\delta_{R}\delta_{R}}^{approx}=\lambda_{\delta_{R}\delta_{R}\delta_{R}}+ \frac{\left(2 \alpha _3^2+16 \rho _2^2+3 \rho _3^2\right) v_R}{24 \sqrt{2} \pi ^2}.\label{ddd_aprox}
\end{align}
In Fig.~\ref{quantum_cor_fig} we show the deviation from the expressions~\eqref{hhh_aprox}-\eqref{ddd_aprox} of the the full quantum corrections due to the entire scalar sector. More precisely we plot
\begin{equation}
|(\lambda_{i,j,k}^{\text{total}})-\lambda_{i,j,k}^{\text{approx}})/\lambda_{i,j,k}^{\text{approx}}|,\label{loop_deviation}
\end{equation}
in the $(m_{\delta_R},\theta)$ plane, where $\lambda_{i,j,k}^{\text{total}}$ are the trilinear couplings with the full quantum corrections included and the indices $\{i,j,k\}$ range on $h$ and $\delta_R$.

Notice that  $\lambda^{approx}_{hhh}$ would be affected by the further quantum corrections in the presence of non-vanishing mixing, as shown in Fig.~\ref{quantum_cor_fig}. Therefore a larger SM deviation than the one in~\eqref{deviation} may result in some portions of the parameter space. This can be understood by noticing that for non zero mixing,  $\lambda_\Phi$ has to be larger than its SM value (see first line in Tab.~\ref{tab:1}), thus affecting the tree level values for the couplings entering directly in the loops. Furthermore, there are contributions depending on both $\alpha_3$ and $\theta$. This is particularly true for the effective $\lambda_{h\delta_{R}\delta_{R}}$, as clear from Fig.~\ref{quantum_cor_fig}, which receives contributions $\sim\alpha_3^2 \theta$. Nevertheless, the approximations in~\eqref{hhh_aprox}-\eqref{ddd_aprox} work quite well for wide regions in the $(m_{\delta_R},\theta)$ plane.

In the natural case with $m_{\delta_{R}} \simeq v_R$ and $\theta$ negligible, in accord with the perturbativity constraints~\cite{Maiezza:2016bzp}, the effective vertices discussed above assume a simple form given in the Appendix~\ref{appendixC}.
In the Fig.~\ref{quantum_cor_fig}, for the case $M_{W_{R}}=6$ TeV we have used  the bounds in~\eqref{mDL} and~\eqref{mDR} on $\rho_2,\rho_3$, while for $M_{W_{R}}=20$ TeV the experimental constraints~\cite{ATLAS:2014kca} on $m_{\delta_{R}}^{++},m_{\delta_{L}}^{++}$ of a few hundred GeV. Assuming larger values, especially in the latter case, changes only slightly the effective vertices and the explicit check shows that the Fig.~\ref{quantum_cor_fig} remains quite stable.

\section{Conclusions and Outlook}\label{V}

The LR symmetric theory has gone through a revival of interest in recent years, and for good reasons. Due to the theoretical limits on its scale, obtained in the early eighties, one had to wait for the LHC in order to hope for its verification. The possible LHC signatures are remarkable:  lepton number violation through the production of same sign charged lepton pairs and the way of directly testing the Majorana nature of RH neutrinos~\cite{Keung:1983uu}. This is connected with the low energy lepton number and lepton flavor violating processes~\cite{Tello:2010am}. Moreover, the theory allows for a direct probe of the origin of neutrino mass and a disentangling of the seesaw mechanism~\cite{Nemevsek:2012iq}, as long as one can measure the masses and mixings of the RH neutrinos~\cite{Das:2012ii, AguilarSaavedra:2012gf} and their Majorana nature~\cite{Gluza:2016qqv}. In recent years, one has also finally computed analytically the RH quark mixing matrix~\cite{Senjanovic:2014pva}.

One cannot overemphasize the importance of the study of the Higgs sector of the theory, especially today when it appears that the SM Higgs boson has been found. The rich scalar sector of the LRSM merges two milestones of the present day phenomenology, the Higgs boson physics and the origin of neutrino mass. An example of the related literature can be found in~\cite{Maiezza:2015lza}, where a LNV Higgs decay is analyzed in the light of LHC. Nevertheless, a complete analysis of the whole phenomenologically relevant parametric space of the scalar potential was still missing, and in the present work we have attempted to fill the gap. In particular we have discussed the scalar spectrum with a generic $t_\beta$ and the spontaneous CP-phase. This configuration, moreover, would be the one needed for a RH scale in the reach of LHC~\cite{Maiezza:2010ic} if the LR symmetry were $\mathcal{P}$. In any case the full knowledge of the scalar masses is fundamental for the matching of the parameters of the model with the relevant physical observables. Such an example is the evolution of the quartics under the RGE's, which requires a direct match with the analytical expression of the masses, in order to ensure the stability of the potential.

We have examined the behavior of the model in three different regime: LHC energy reach, next 100 TeV hadronic collider and very high energy, in accordance with the $SO(10)$ GUT constraints.
In the first regime, our analysis shows that the model lives dangerously. While it is not ruled out from LHC reach, new physics beyond the LRSM is already required at energy $\sim10 M_{W_R}$. This cut-off implies stringent bounds for the entire scalar spectrum, and except that for $m_{\delta_{R}}$ that might be light as an effective SM singlet, all other states end up too heavy to be seen at the LHC. A light $\delta_R$ could have direct implications for the standard-like Higgs physics, with fairly large deviations of the Higgs self-couplings from the SM predictions, measurable in the near future.

The second energy regime considered is the one of next hadronic collider. Here the model becomes more natural. The cut-off appears far away from $M_{W_{R}}$, although well below GUT scale.

In the last energy regime we discussed the $SO(10)$ embedding, within the scenario of two-step symmetry breaking. We have shown that the usual picture fits well with the RGE evolution of the whole parametric space of the LRSM, as long as the quartics are fairly small of order of 10$^{-1}$.

We conclude with the vertex renormalization for the phenomenologically important system $h,\delta_R$, showing the anatomy of the quantum corrections that may be dominant in some regions of the parametric space.

\section*{Acknowledgments}

We thanks Fabrizio Nesti and Vladimir Tello for useful discussions. JV  was funded by Conicyt PIA/Basal FB0821 and Fondecyt project  N. 3170154.

 \appendix
\section{Neutral scalar masses}\label{appendixA}

Here we discuss the neutral mass matrix for the scalar potential in~\eqref{CV}. What in principle could be a complicated $4 \times 4$ matrix, reduces effectively to the $\{h_{SM}, \delta\}$ system, since the flavor violating neutral components H and A decouple and form a part of the super-heavy doublet $\phi_{FV}$ with the mass $m^2_{\phi_{FV}}=\frac{\alpha_3 }{c_{2\beta}} v_R^2$.

Some comments are in order. First of all, the mass of the heavy doublet $\phi_{FV}$ receives corrections of the order $v^2$ that we discard because of the strong limit on its mass of around 20 TeV~\cite{Bertolini:2014sua} and
the $\phi_{FV}$ components (scalar and pseudo-scalar) are degenerate for any phenomenological purpose.
For the same reason we neglect in $m_{\delta_R}^2$ in Tab.~\ref{tab:1} those terms suppressed as $1/m_{FV}^2$ and moreover, we neglect the small mixing between $\delta$ and $\phi_{FV}$ states, which can be relevant in the case of their quasi degeneracy, of little phenomenological interest, in which case one could trust the tree-level anyway. It is worth noticing that a very light $\delta_R$, well below the electro-weak scale, requires some more care because of potential FCNC effects. This subject has been recently studied in~\cite{Dev:2016vle}, in which a strong constraint on $\theta$ is obtained. However, this does not affect our results, since we consider $m_{\delta_R}\geq m_h$. In such a case, this mixing is suppressed by the electro-weak scale, completely negligible due to the huge mass of $\phi_{FV}$ field. The only mixing to consider is between $h_{SM}$ and $\delta$, and only if $\delta$ is relatively light.

The mass matrix for the $\{h_{SM}, \delta\}$ system is then found to be
\begin{equation}
	M_0^2 \simeq \left[ \begin {array}{cccc}
\noalign{\medskip}
 m_{h_{SM}}^2 & m^2_{\delta h}\\
\noalign{\medskip}
m^2_{\delta h}  & m_{\delta}^2\\
\end {array} \right], \label{completemassmatrix}
\end{equation}
where
\begin{align}
 m^2_{h_{SM}}&=4\lambda_{\Phi} v^2,\\
 m_{\delta}^2& = 4\,\rho_{{1}}v_R^2,  \\
 m^2_{\delta h}& =2\alpha v v_R,
\end{align}
and $\lambda_{\Phi}$ and $\alpha$ are given by~\eqref{l_a}.

This matrix has the following eigenvalues (it is effectively the SM augmented by a real scalar singlet studied in~\cite{Gupta:2013zza})
\begin{align}\label{Higgsmass}
m_{h,\delta_R}^2 = & 2 \left[\rho _1 v_R^2+v^2 \lambda _{\Phi }\mp\left(\rho _1 v_R^2-v^2 \lambda _{\Phi }\right)\times \right. \\ \nonumber
& \left. \sqrt{\frac{\alpha ^2 v^2 v_R^2}{\left(\rho _1 v_R^2-\lambda_\Phi  v^2\right)^2}+1}\,\, \right]
\end{align}
where $h = c_\theta h_{SM} - s_\theta \delta, \,\,\delta_R = s_\theta h_{SM} + c_\theta \delta$, with the mixing given by
\begin{equation}
t_{2 \theta} = \frac{\alpha v v_R}{\rho_1 v_R^2-\lambda_\Phi v^2}\,.
\end{equation}
Finally we quote the expressions of $\lambda_\Phi,\rho_1,\alpha$ in terms of the masses $m_h,m_{\delta_R}$ and the mixing~\cite{Gupta:2013zza}
\begin{align}
 \lambda _{\Phi }&=\frac{c_{2\theta} \left(m_h^2-m_{\delta_R}^2\right)+m_{\delta_R}^2+m_h^2}{8 v^2}\,, \\
\rho _1&=\frac{c_{2\theta} \left(m_{\delta_R}^2-m_h^2\right)+m_{\delta_R}^2+m_h^2}{8 v_R^2}\,, \\
\alpha& =\frac{s_{2\theta }\left(m_{\delta_R}^2-m_h^2\right)}{4 v v_R}\,.
\end{align}
%

\section{A look at RGE's at higher order}\label{appendixB}

In this appendix, we estimate the impact of the two-loop corrections to the running of the quartic couplings in the potential. Since the complete two-loop corrections is out of the scope of this work, we consider the corrections due to the scalar self-couplings only. We expect that the leading contribution
from the two-loop is due to the self-quartics part, in full analogy with the one-loop result~\cite{Rothstein:1990qx} where the full expressions are provided. Also, the gauge couplings remain always smaller than unity, even for larger quartics (this is precisely the case in which two-loop might be relevant), and for this reason they play a secondary role.
With this in mind and as an illustrative example, we show the partial two-loop and one-loop $\beta$-function of $\lambda_1$ for a direct comparison, including only the contributions of the scalar quartics since only these may become  dangerously large
\begin{widetext}
\begin{eqnarray}\label{1_2_match}
(4\pi)^2 \beta_{(1-loop)}^{\lambda_1} &\supset & 6 \alpha _1^2+6 \alpha _3 \alpha _1+2.5 \alpha _3^2+32 \lambda _1^2+64 \lambda _2^2+16 \lambda _3^2+48 \lambda _4^2+16 \lambda _1 \lambda _3 \,; \\ \nonumber
(4\pi)^2 \beta_{(2-loop)}^{\lambda_1} &\supset & \frac{1}{384 \pi^2}\left\{-36 \alpha _1^2 \left(\alpha_3-30 \lambda_1\right)-2 \alpha_1 \left[\alpha_3 \left(19 \alpha_3-540 \lambda _1\right)+48 \alpha_2 \left(\alpha_2+3 \lambda_4\right)\right]+826 \alpha_3^2 \lambda_1 \right. \\ \nonumber
&-& 48 \alpha_2^2 \left(\alpha_3-94 \lambda_1+8 \lambda_2+4 \lambda _3\right)-144 \alpha_2 \alpha_3 \lambda_4-24 \alpha_1^3-13 \alpha_3^3+2304 \lambda_1 \rho_1^2 +3456 \lambda_1 \rho_2^2 \\ \nonumber
&+& 432 \lambda_1 \rho_3^2 + 2304 \lambda_1 \lambda_4^2 + 3456 \lambda_1 \rho _4^2+2304 \lambda_1 \rho_1 \rho_2+1424 \lambda_1^3-384 \lambda_3^3+14592 \lambda_1 \lambda_2^2\\ \nonumber
&+& 2304 \lambda_1 \lambda_3^2 - \left.3328 \lambda_2 \lambda_4^2-1792 \lambda_3 \lambda_4^2+1152 \lambda_1^2 \lambda_3-5632 \lambda_2^2 \lambda_3\right\}. \nonumber
\end{eqnarray}
\end{widetext}
Let us emphasize once again that in section~\ref{IV} the complete one-loop RGE's were used. The expressions in~\eqref{1_2_match} can be worked out from the general formalism in~\cite{Machacek:1984zw} and are both normalized with $(4\pi)^2$ for a direct comparison. A drastic gap between the size of the coefficients of one-loop and two-loops is evident, although the number of the contributions clearly increases for the latter. Similar expressions hold for the other quartics.

\section{Effective trilinear vertices}\label{appendixC}

Here we report the expressions of the trilinear vertices with negligible mixing $\theta$
\begin{widetext}
\begin{align}
\lambda_{hhh}^{eff.}&=\sqrt{2} \lambda_{\Phi}v+\frac{1}{384 \pi ^2 \alpha _3 \rho _2 \left(2 \rho_1-\rho _3\right) v_R^2}\left(432 \sqrt{2} \alpha _3 \lambda_{\Phi}^2 \rho _2 \rho_1 v v_R^2-216 \sqrt{2} \alpha _3 \lambda_{\Phi}^2 \rho _2 \rho _3 v v_R^2-9 \sqrt{2} \alpha _3^4 \rho _2 v^3 \right.\\ \nonumber
&\hspace{3 em}\left.-2 \sqrt{2} \alpha _3^4 \rho _3 v^3+4 \sqrt{2} \alpha _3^4 \rho_1 v^3+256 \sqrt{2} \lambda_{\Phi}^3 \rho _2 \rho_1 v^3-128 \sqrt{2} \lambda_{\Phi}^3 \rho _2 \rho _3 v^3 \right)\,, \\
\lambda_{hh\delta_{R}}^{eff.}&=\frac{v^2 \left(\alpha _3^2 \left(8 \rho _2 \rho_1-2 \rho _3 \rho_1-9 \rho _2 \rho _3+4 \rho_1^2\right)+32 \lambda_{\Phi}^2 \rho _2 \left(2 \rho_1-\rho _3\right)\right)}{32 \sqrt{2} \pi ^2 \rho _2 \left(2 \rho_1-\rho _3\right) v_R}\,,\\
\lambda_{h\delta_{R}\delta_{R}}^{eff.}&=\frac{\alpha _3 v \left(-2 \rho _3 \left(4 \lambda_{\Phi}\rho _2+\left(2 \rho _2+\rho_1\right){}^2\right)+4 \rho_1 \left(4 \lambda_{\Phi}\rho _2+\left(2 \rho _2+\rho_1\right){}^2\right)-3 \rho _2 \rho _3^2\right)}{16 \sqrt{2} \pi ^2 \rho _2 \left(2 \rho_1-\rho _3\right)}\,,\\
 \lambda_{\delta_R\delta_R\delta_R}^{eff.}&=\sqrt{2} \rho _1 v_R+\frac{1}{48 \pi ^2 \rho _2 \left(2 \rho_1-\rho _3\right)}\left(4 \sqrt{2} \alpha _3^2 \rho _2 \rho_1 v_R-2 \sqrt{2} \alpha _3^2 \rho _2 \rho _3 v_R+78 \sqrt{2} \rho _2 \rho_1^3 v_R-2 \sqrt{2} \rho _3 \rho_1^3 v_R\right.\\ \nonumber
&\left.+48 \sqrt{2} \rho _2^2 \rho_1^2 v_R-39 \sqrt{2} \rho _2 \rho _3 \rho_1^2 v_R+32 \sqrt{2} \rho _2^3 \rho_1 v_R-24 \sqrt{2} \rho _2^2 \rho _3 \rho_1 v_R-3 \sqrt{2} \rho _2 \rho _3^3 v_R-16 \sqrt{2} \rho _2^3 \rho _3 v_R+4 \sqrt{2} \rho_1^4 v_R    \right)\,.
\end{align}
\end{widetext}
%

\noindent

\vskip 2 cm

\def\arxiv#1[#2]{\href{http://arxiv.org/abs/#1}{[#2]}}

\end{document}